\documentclass[preprint2]{aastex}

\usepackage[caption=false]{subfig}
\usepackage{graphicx}
\usepackage{epstopdf}
\usepackage{amsmath}
\usepackage{longtable}
\usepackage{lscape}
\usepackage{natbib}
\usepackage{url}
\usepackage{color}

\shorttitle{A new catalog of variable stars in the M37}
\shortauthors{Chang et al.}

\begin{document}
\captionsetup[subfigure]{labelformat=empty}

\title{A New Catalog of Variable Stars in the Field of the Open Cluster M37}

\author{S.-W. Chang\altaffilmark{1}, Y.-I. Byun\altaffilmark{2}, and J. D. Hartman\altaffilmark{3}}

\altaffiltext{1}{Institute of Earth$\cdot$Atmosphere$\cdot$Astronomy, Yonsei University, Seoul 120-749, South Korea; seowony@galaxy.yonsei.ac.kr}
\altaffiltext{2}{Department of Astronomy and University Observatory, Yonsei University, Seoul 120-749, South Korea; ybyun@yonsei.ac.kr}
\altaffiltext{3}{Department of Astrophysical Sciences, Princeton University, Princeton, NJ 08544, USA}

% Abstract
\begin{abstract}
We present a comprehensive re-analysis of stellar photometric variability in the field of the open cluster M37 following the application of a new photometry and de-trending method to MMT/Megacam image archive.  This new analysis allows a rare opportunity to explore photometric variability over a broad range of time-scales, from minutes to a month.  The intent of this work is to examine the entire sample of over 30,000 objects for periodic, aperiodic, and sporadic behaviors in their light curves.  We show a modified version of the fast $\chi^{2}$ periodogram algorithm (F$\chi^{2}$) and change-point analysis (CPA) as tools for detecting and assessing the significance of periodic and non-periodic variations.  The benefits of our new photometry and analysis methods are evident.  A total of 2306 stars exhibit convincing variations that are induced by flares, pulsations, eclipses, starspots, and unknown causes in some cases.  This represents a 60\% increase in the number of variables known in this field.  Moreover, 30 of the previously identified variables are found to be false positives resulting from time-dependent systematic effects.  New catalog includes 61 eclipsing binary systems, 92 multiperiodic variable stars, 132 aperiodic variables, and 436 flare stars, as well as several hundreds of rotating variables.   Based on extended and improved catalog of variables, we investigate the basic properties (e.g., period, amplitude, type) of all variables.  The catalog can be accessed through the web interface (\url{http://stardb.yonsei.ac.kr/}). 
\end{abstract}
\keywords{binaries: eclipsing --- catalogs --- open clusters and associations: individual (M37) --- stars: flare --- stars: oscillations --- stars: variables: general}

% Introduction
\section{Introduction}
The rich open cluster M37 (NGC 2099) in the constellation Auriga has been photometrically monitored to search for new variables with different motivations \citep{kis01,kan07,mes08,har08a}.  The combination of short and deep exposures with similar sky coverage allowed time-series photometry of hundreds to thousands of stars from the brightest to the dimmest within the cluster field ($10 < V < 23$).  Their studies indicate that the fraction of variable sources $f_\mathrm{var}$ increases as photometric precision of the survey gets better.

\citet{kis01} performed the first variability survey for the central ($29\arcmin\times18\arcmin$) field of the M37 with a $R_{C}$-filter photometry and found 7 variable stars among 2300 stars ($f_\mathrm{var}\sim0.3\%$).  Due to poor photometric quality, their variables were with large variability and included three W UMa-type eclipsing binaries (EBs), two high-amplitude pulsating stars, and two long-period detached EB candidates. 

\citet{kan07} and \citet{mes08} used the same data set to search for short-period pulsating variables (e.g., $\delta$ Scuti- and $\gamma$ Doradus-type stars) and rotating variables of later spectral types (F--K), respectively.  Using $V$-band time-series observations, these two studies found the total of 153 variable stars among 12,000 stars ($f_\mathrm{var}\sim1.3\%$), including seven previously known variables, in the field of the cluster ($22.2\arcmin\times22.2\arcmin$).
  
Lastly, the most extensive variability studies in this field were carried out by \citet{har08b,har09a,har09b} as the by-products of a deep survey for transiting planets in the M37 field ($24\arcmin\times24\arcmin$) with the 6.5m MMT telescope.  The number of known variable stars increased by a factor of $\sim$10.  By applying several statistical algorithms, they identified 1445 variables among 23,000 stars ($f_\mathrm{var}\sim6.3\%$), of which 99\% were new discoveries \citep{har08b}.  This catalog of variable stars provides the largest homogeneous set of stellar rotation periods for a cluster older than 500 Myr, which showed the crucial role of the rotation in the stellar angular momentum evolution \citep{har09a}. 

In this paper, we search new variable phenomena from the same data set of \citet{har08a} with vastly improved photometric method (\citealt{cha15}, hereafter Paper I).  In Section 2, we give a brief description of new light curves and its temporal properties.  We present a comprehensive re-analysis of photometric variability in Section 3.  In Section 4, we present a new catalog of variable stars and various variability characteristics.  We compare our findings with previous works and discuss the nature of each variability class separately in Section 5.  

% Figure 1
\begin{figure}[t]
\centering
  \includegraphics[width=0.5\textwidth, angle=0]{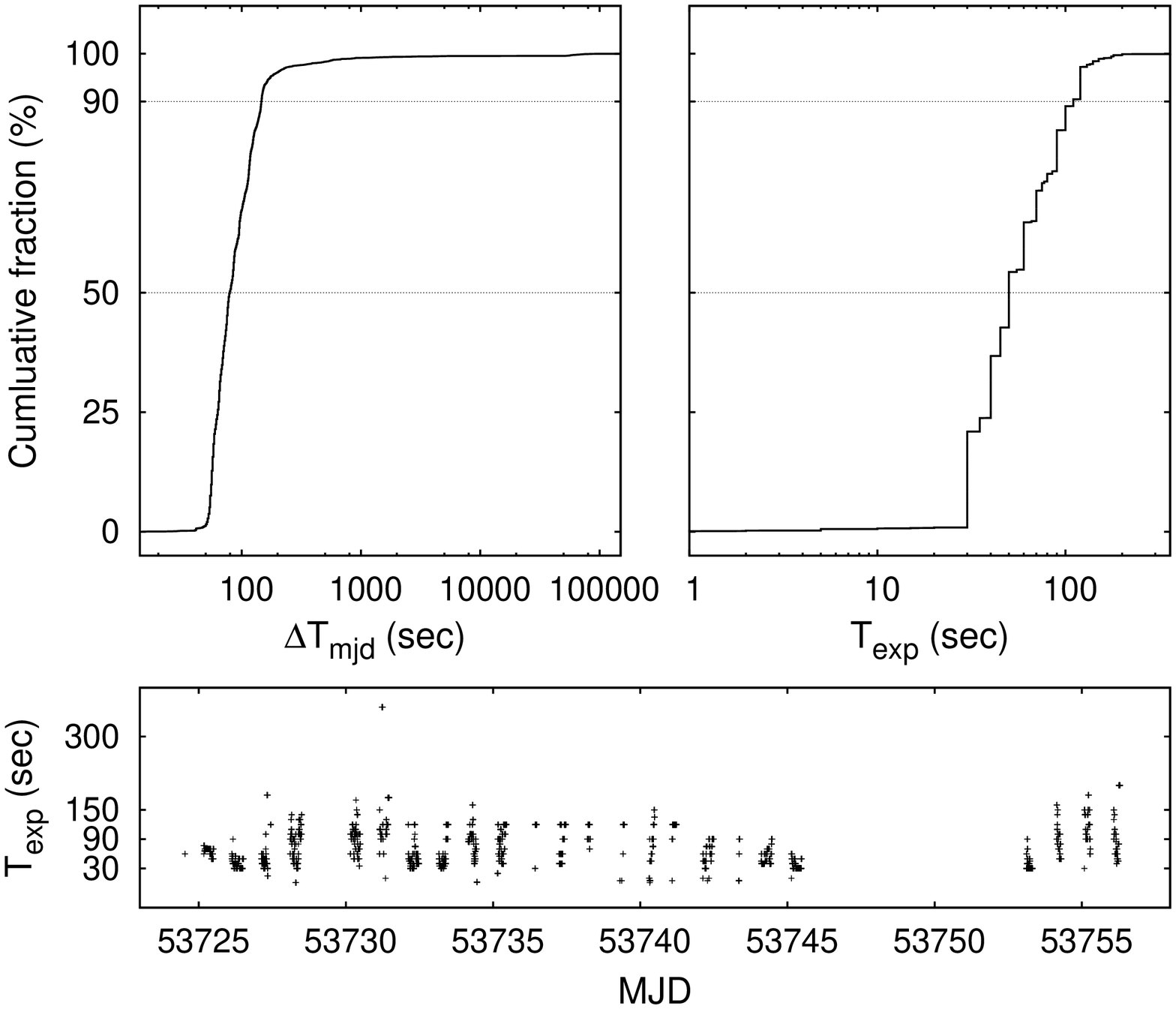}    
  \caption{Observational cadence of the MMT transit survey, showing cumulative fraction function of time intervals $\Delta T_\mathrm{mjd}$ between two consecutive frames (top left) and the exposure times $T_\mathrm{exp}$ (top right).  We use logarithmic scale for the horizontal axis in these two panels.  The bottom panel is the distribution of exposure times as a function of the modified Julian date (MJD).}
  \label{Fig1}
\end{figure}     

% Section 2
\section{Data description}
We refer the reader to Paper I for a detailed description of our photometric reduction and light curve production applied to the archival imaging data of MMT transit survey.  Here we provide only a brief summary of the key features.  In Paper I, we introduced new methods for robust high-precision photometry from well-sampled images of a non-crowded field with strongly varying PSF.  The archival light curves from the original image subtraction procedure exhibit many unusual outliers, and more than 20\% of data get rejected by the simple filtering algorithm adopted by early analysis.  We note that some of these outliers are results of intrinsic variability of short time scales.  In order to achieve better photometric precisions and also to utilize all available data, the entire imaging database were re-analyzed with our multi-aperture indexing photometry and a set of sophisticated calibration procedures.  This algorithm finds an optimal aperture for each star with a maximum signal-to-noise ratio, and also treat peculiar situations where photometry returns misleading information with more optimal photometric index.   We also applied photometric de-trending algorithm based on a hierarchical clustering method.  Our method removes systematic variations that are shared by light curves of nearby stars while true variabilities are preserved.  Consequently, our method utilizes nearly 100\% of available data and reduce the RMS scatter several times smaller than original analysis for brighter stars.

Our new data set includes 30,294 objects, including both point and extended sources.  For blended targets, the original image subtraction light curve is still of higher precision than the present aperture photometry light curves, and thus for these objects the archival light curve is used.  In order to pick out such cases, we utilized a simple method for quantifying the level of blending (see \citealt{irw07}).
\begin{displaymath}
b = \frac{\chi^{2}-\chi_{poly}^{2}}{\chi^{2}}
\end{displaymath}
where
\begin{equation}
\chi^{2} = \sum^{}_{p}\frac{(\sum m_{p}-\bar{m}_{j})^{2}}{\sigma_{p}^{2}}
\end{equation} for light curve points $m_{p}$ with uncertainties $\sigma_{p}$, and $\chi_{poly}^{2}$ is the same statistic measured with respect to a 4th order polynomial fit.  We adopt the value $b > 0.6$ for selecting significant blended light curves, and then count the number of blended ones (3746 objects).  

Figure \ref{Fig1} shows that the cumulative fraction function of temporal intervals $\Delta T_{mjd}$ between two consecutive frames and the exposure times $T_{exp}$.  This observing cadence is sensitive to variability on timescales ranging from tens of seconds (30 seconds) to one month ($\sim$31.7 days).  In particular, the time resolution allows an analysis of the short-term variability, which is less than $\sim$80 seconds over 50\% of the data and $\sim$150 seconds over 90\% of the data.  

% Figure2
\begin{figure*}[t]
\centering
  \subfloat[]{\includegraphics[width=0.5\linewidth, angle=0]{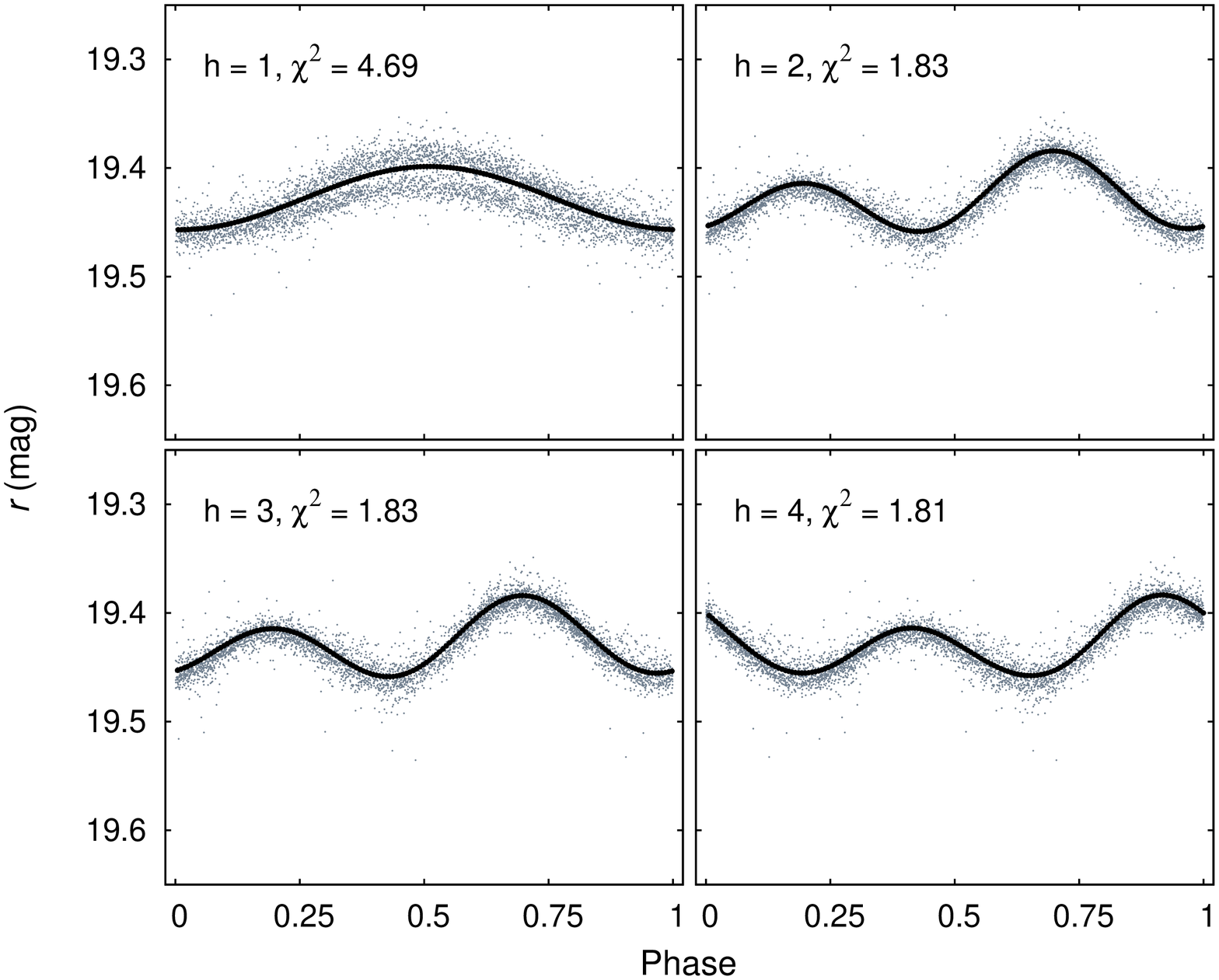}}
  \subfloat[]{\includegraphics[width=0.5\linewidth, angle=0]{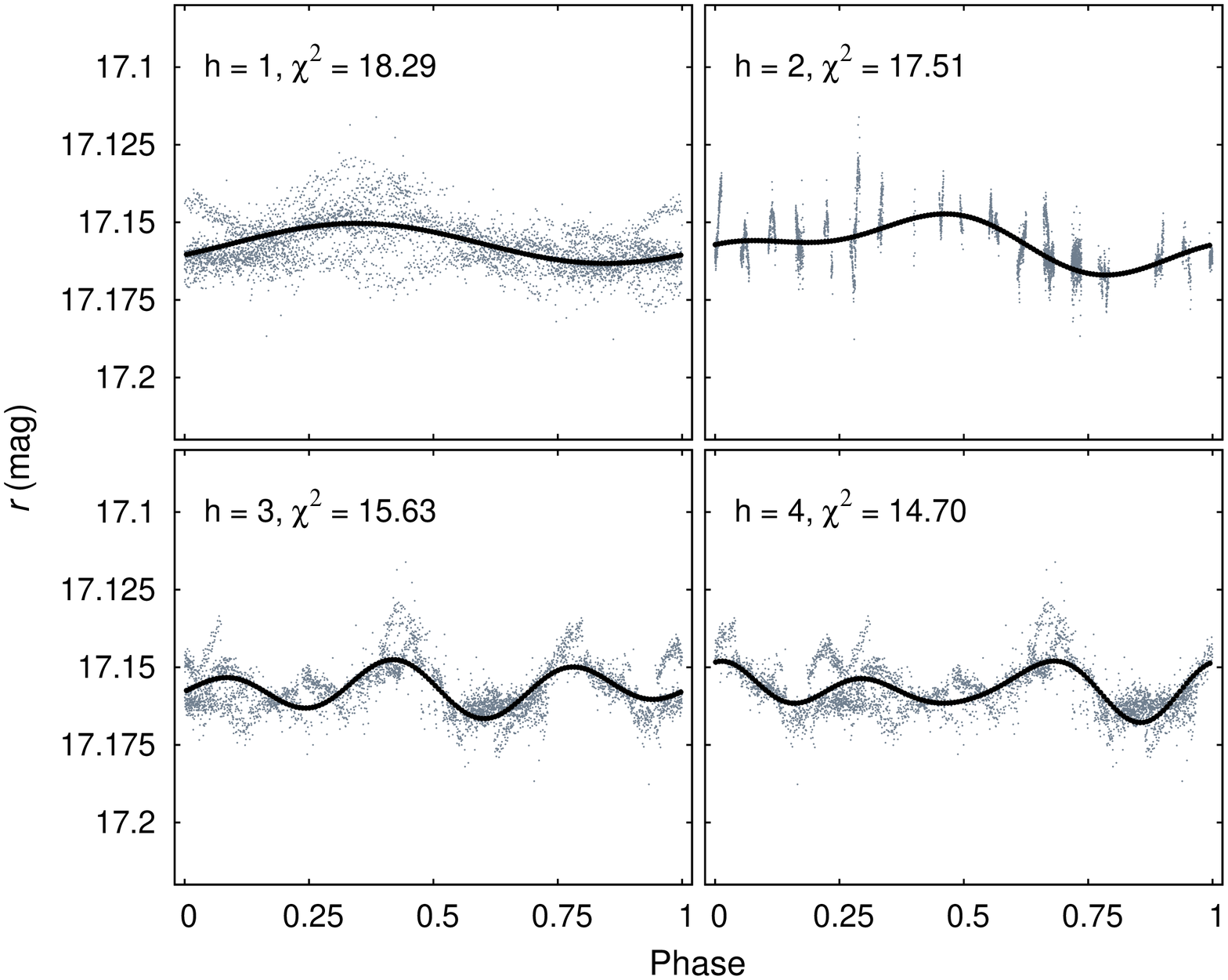}}\vspace*{-1.3em}
  \caption{Example light curves of a periodic star (four plots at left panel) and another quasi-periodic star (four plots at right panel) with different periods detected by the first feature detection algorithm, respectively.  The shape of periodic feature is varied widely by the selected harmonic model $h$ (from 1 to 4), but we can choose the best candidate with $\chi^{2}$ statistics.}
  \label{Fig2}
\end{figure*}

% Section 3
\section{Variability-analysis methods}
Variability features can be divided into two broad categories: periodic and non-periodic feature.  The former is easier to detect and model using statistical and model-specific fitting procedures as a form of Fourier series (e.g., \citealt{deb07,deb09,ric11}).  But the latter is rather ambiguous, since it is not always possible to characterize flux variation and to define a reliable range for some parameters.  To overcome this difficulty, \citet{ric11} calculated many basic statistics that can describe the distribution of fluxes even in the limit of few data points.  One interesting approach is to select stochastically varying features in quasar-like sources.  The correlated variability of quasars is well parameterized by a damped random walk model, providing a simple statistical description of their characteristic timescales and amplitudes (e.g., \citealt{kel09,but11}).  In this work we use a modified version of the fast $\chi^{2}$ periodogram algorithm (F$\chi^{2}$) and the change-point analysis (CPA) method for detecting and assessing the significance of periodic and non-periodic signals, respectively.

% Section 3.1
\subsection{Generic periodic feature detection}
We first searched a periodic feature by F$\chi^{2}$ technique that calculates the minimized $\chi^{2}$ as a function of frequency at the desired number of harmonics \citep{pal09}.  With the best-fit frequency, we compared the ratio between the variance of the data after subtracting the best-fit model and full-amplitude of model function.  This algorithm use a Fourier series truncated at harmonic $h$ as a model function:
\begin{equation}
\Phi_{h} = A_{0} + \sum_{h=1}^{\mathrm{4}} A_{2h-1} \sin(h 2 \pi f t) + A_{2h} \cos(h 2\pi f t) 
\end{equation}  The fitting coefficients ($A_{0}$, $A_{2h}$, and $A_{2h-1}$) obtained in each harmonic model provide a reasonable description of a generic periodic feature, irrespective of the feature shape, as discussed by \citet{deb09}.  Figure \ref{Fig2} shows example light curves of both periodic (new variable star; V1983) and long-term (V1092) variable stars in our sample, respectively.  Because the optimal choice of $h$ depends on the true shape of the signal, the fit quality is evaluated for each model with standard $\chi^{2}$ statistics, and then checked the residual (phase-folded) light curves after the subtraction of the best-fit model.  To test the significance of a signal, we use the ratio between the full-amplitude of the model function ($\mathrm{Famp}$ = $m_{max} - m_{min}$) and the RMS scatter ($\sigma_{a}$) of the data after model subtraction:
\begin{equation}
\mathrm{Famp} \ge \lambda \sigma_{a}.
\end{equation} The lower limit for the ratio is empirically set to $\lambda$ = 0.95 based on light curve inspections.  $\lambda$-values less than 0.95 indicate that we do not expect any significant (semi-) periodic signals for a given light curve.  Even when the source is poorly described by the harmonic fit, this indicator provides a hint of a generic periodic signal (see right panel of Figure \ref{Fig2}).

% Figure 3
\begin{figure*}[t]
\centering
  \subfloat[]{\includegraphics[width=0.5\linewidth, angle=0]{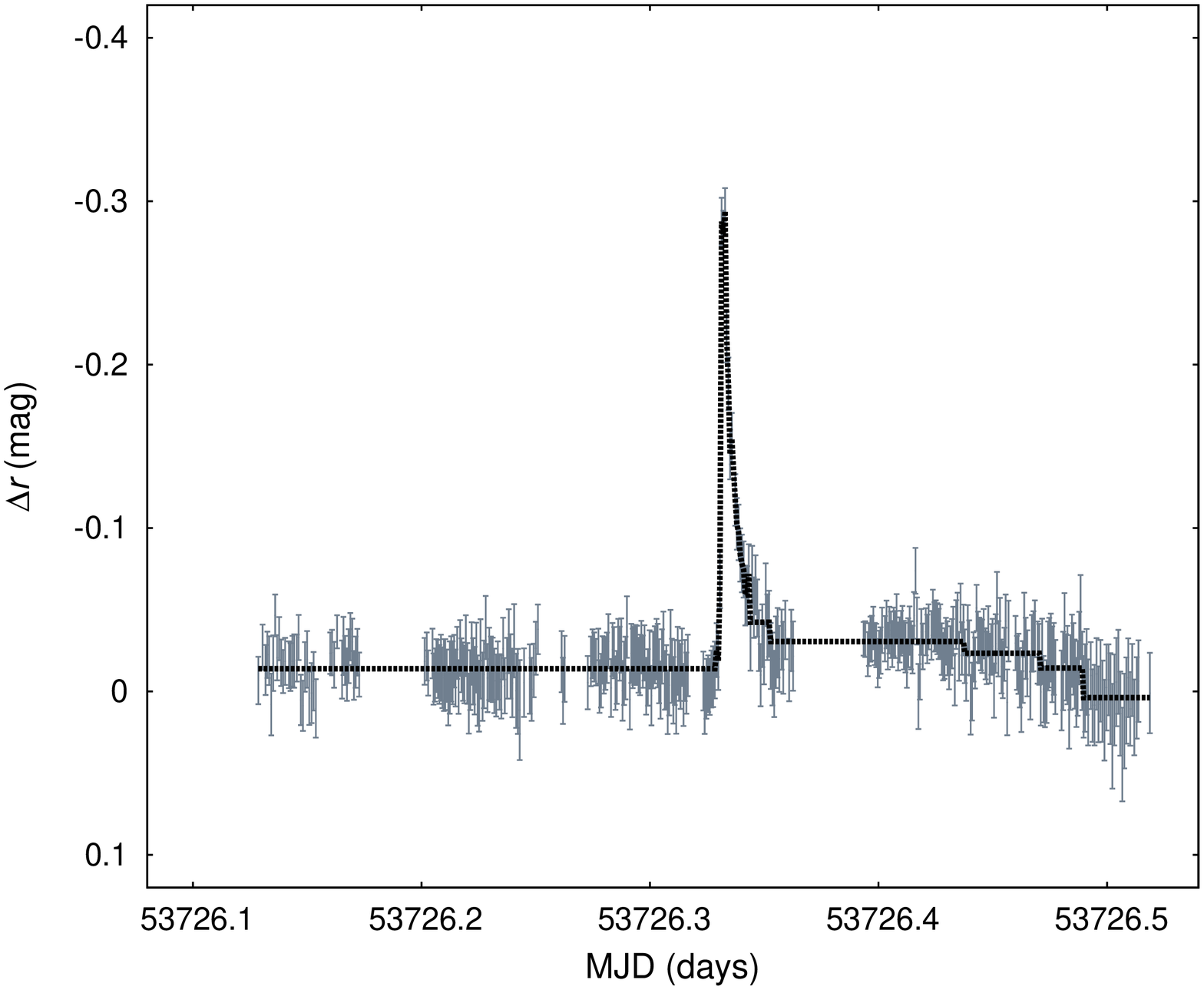}}
  \subfloat[]{\includegraphics[width=0.5\linewidth, angle=0]{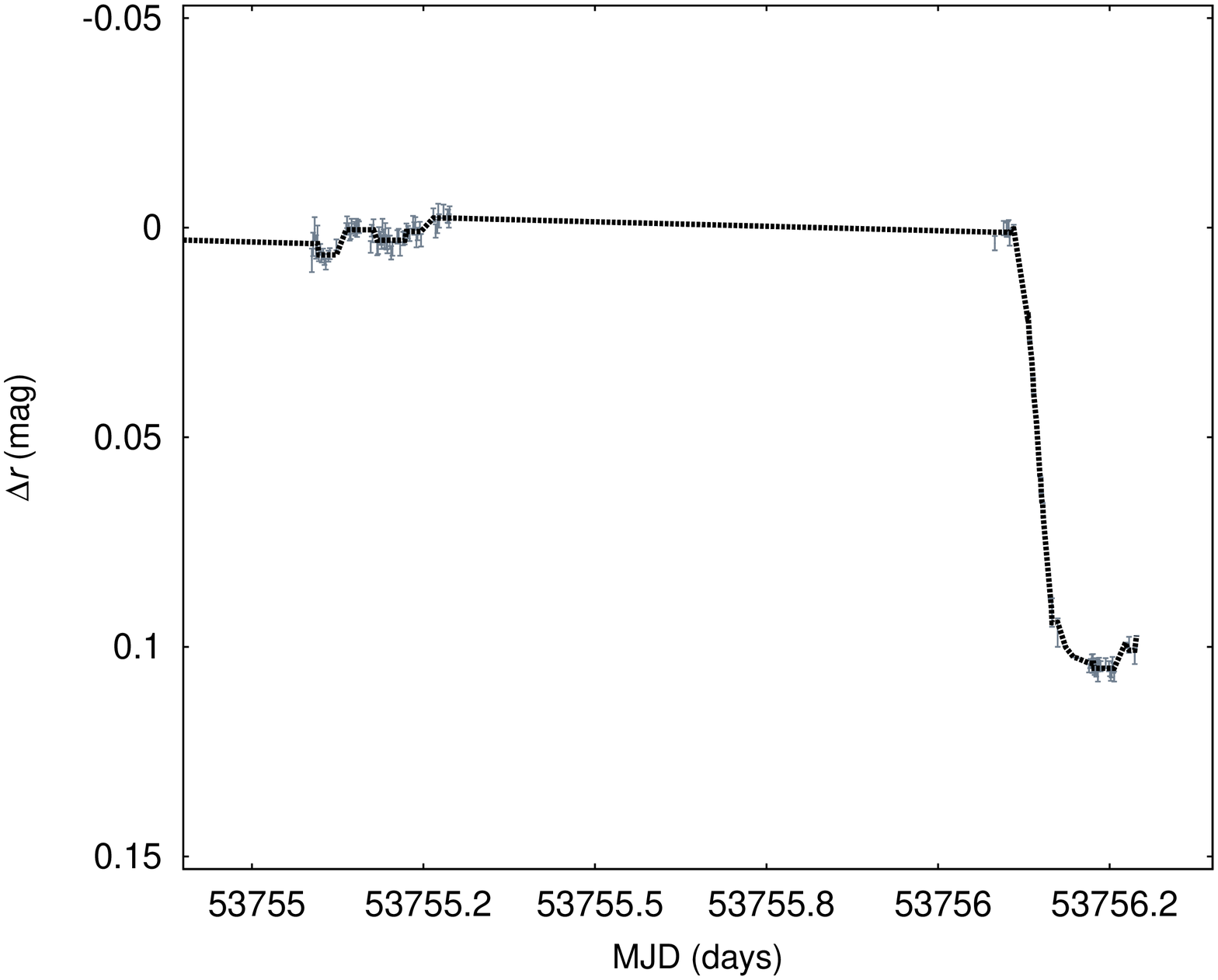}}\vspace*{-1.3em}
  \caption{Example light curves of flare-like (left panel) and eclipsing-like (right panel) events detected by the second feature detection algorithm.  We used the default values of the variable parameters ($N_{1}$=3, $N_{2}$=1, $N_{3}$=3) as the detection thresholds.  The CPA model is over-plotted in dashed lines.}
  \label{Fig3}
\end{figure*}

% Section 3.2
\subsection{Arbitrary-shaped and sporadic event detection}
\subsubsection{Change-point analysis (CPA)}
Most variability detection methods require conventional models that are mainly focused on the periodic signals, and are not suitable for the study of arbitrary-shaped, non-periodic, and sporadic variations.  Also, in many cases, signal estimation is equated with smoothing of data for de-noising.  This sometimes discards vital information in time series data.  We introduce a non-parametric method to extract all significant features based on the CPA using local statistics.

Change-point analysis is a method to identify abrupt variations in sequential data.  It is widely used in the statistics and data mining communities as well as in the field of time domain astronomy (see e.g., \citealt{sch11} and reference therein).  Using a combination of cumulative sum scheme (CUSUM) and bootstrap rank statistics \citep{tay00}, we produce a series of estimated change-points which correspond to the moments of apparent systematic changes.  For a given dataset \{$x_{1}$, $x_{2}$, $\cdots$, $x_{n}$\} of size $n$, the CUSUM values are given by 
\begin{equation}
S_{t} = \sum_{i=1}^{t} (x_{i} - \bar{x})
\end{equation} for $t$ = $0, 1, \cdots, n$, where $S_{0}$ = 0 and the mean value $\bar{x}$ for the sample of $n$ values of $x_{i}$.  An inflection point at which the sign of the CUSUM slope changes is used to determine whether a given interval of data should be kept as one ($\bar{x}_1$=$\bar{x}_2$=$\cdots$=$\bar{x}_n$=$\bar{x}$) or subdivided into two subintervals ($\bar{x}_1$=$\cdots$=$\bar{x}_{p}$$\neq$$\bar{x}_{p+1}$=$\bar{x}_n$).  Based on $N(\ge1,000)$ bootstrap samples that are randomly re-ordered original values, we estimate the confidence level (c.l.) to reduce the false positives by random noise.  Here are the main procedure of each bootstrap process: 
\begin{itemize}
\item We generate a bootstrap sample of $n$ dataset, denoted \{$x_{1}^{b}$, $x_{2}^{b}$, $\cdots$, $x_{n}^{b}$\}, by randomly re-ordering the original $n$ values without replacement.

\item Based on the bootstrap dataset, we calculate the bootstrap CUSUM $S_{t}^{b}$ by using equation (4), and then obtain the maximum, minimum, and difference of the bootstrap CUSUM.
\begin{equation}
S_{diff}^{b} = S_{max}^{b} - S_{min}^{b},
\end{equation} where
\begin{displaymath}
S_{max}^{b} = \operatorname*{max} \{S_{0}^{b}, S_{1}^{b}, \cdots, S_{n}^{b}\},
\end{displaymath}
\begin{displaymath}
S_{min}^{b} = \operatorname*{min} \{S_{0}^{b}, S_{1}^{b}, \cdots, S_{n}^{b}\}.
\end{displaymath}

\item We check whether the bootstrap difference $S_{diff}^{b}$ is less than the original difference $S_{diff} = S_{max} - S_{min}$.
\end{itemize} It is straightforward to derive estimates of confidence level as follows:
\begin{equation}
\mathrm{c. l.} = 100 \times \frac{\sum X_{b}}{N}\%,
\end{equation} where
\begin{displaymath}
X_{b} = \left\{ 
  \begin{array}{l l}
    1 & \quad \text{if $S_{diff}^{b} < S_{diff}$}\\
    0 & \quad \text{if $S_{diff}^{b} \ge S_{diff}$}
  \end{array} \right.
\end{displaymath}  Typically more than 90\% confidence level is suitable for most cases.  When a sudden change is detected, the location of change-point is initially determined as follow:
\begin{equation}
p_{t} = \operatorname*{arg\, max}_{t\in\left[0,n\right]} \left|S_{t}\right|,
\end{equation} where $p_{t}$ denotes the last point before the change occurred.  The time-series data is split into two segments on each side of the change point, and the analysis is repeated for each segment until no more significant change point is detected.  

Once this set is generated, all the change-point locations and their confidence levels are re-estimated in a backward elimination manner.  Using only the sample in the segment bounded by adjacent change-points, we re-calculate these parameters detected by the forward procedure to obtain more accurate estimations.  When any initial change-point fails the criteria in the backward procedure, we eliminate this change-point estimate.  This process repeats until no further improvement is possible.  The final change points thus define the segments (i.e., piecewise constant level sets), characterized by the start and end time of a given interval with its mean and variance.  After this procedure, signal detection process is performed for each segment separately.  

\subsubsection{Example of flare-like or eclipse-like event detection}
To detect significant outlying features occurring at specific levels in light curve, we adopt a simple criteria similar to Micro-lensing Alert system \citep{gli01} in the presence of hetero-scedastic measurement errors ($w_{i}$):
\begin{mathletters}
\begin{equation}
x_{i} - {\bar{x}_{L}} < 0, 
\end{equation}
\begin{equation}
\frac{\left|x_{i} - {\bar{x}_{L}}\right|}{\sigma_{L}} \ge N_{1},
\end{equation}
\begin{equation}
\frac{\left|x_{i} - {\bar{x}_\mathrm{L}} + w_{i}\right|}{\sigma_{L}} > N_{2},
\end{equation}
\begin{equation}
ConM \ge N_{3},
\end{equation} where the mean $\bar{x}_{L}$ and deviation $\sigma_{L}$ are the local statistics for a given segment, $w_{i}$ is the photometric error at epoch $i$, and $ConM$ is the number of consecutive points which satisfy the equations (8a--8c).  The different combination of $N_{1,2,3}$ values are used for maximize the detection efficiency of significant outlying features.  The values of $N_{1,2,3}$ were taken to be at least larger than 3, 1, and 3, respectively.  As a result, statistically significant events were flagged as candidate flare sources if at least three contiguous points satisfy the above criteria.
\end{mathletters}  Example of flare-like feature detection is shown in the left panel of Figure \ref{Fig3}.  By applying the CPA method to the entire light curves, we efficiently identified several hundred instances of abrupt brightness changes without any smoothing or interpolation of the raw data.

We use the same limits of CPA thresholds to find eclipsing-like event detection, except the first term of equation (8), i.e., $x_{i} - {\bar{x}_{L}} > 0$.  The CPA method is useful to detect the moments of eclipse ingress, center, or egress in cases where the eclipsing pattern is not repeated or the coverage is not sufficient for the detection through conventional period analysis (e.g., Box-fitting Least-Squares method).  Example of eclipsing-like feature detection is shown in the right panel of Figure \ref{Fig3}.  We will discuss this issue further in Section 5.1.  

% Figure 4
\begin{figure*}[t]
\centering
  \subfloat[]{\includegraphics[width=0.34\linewidth, angle=0]{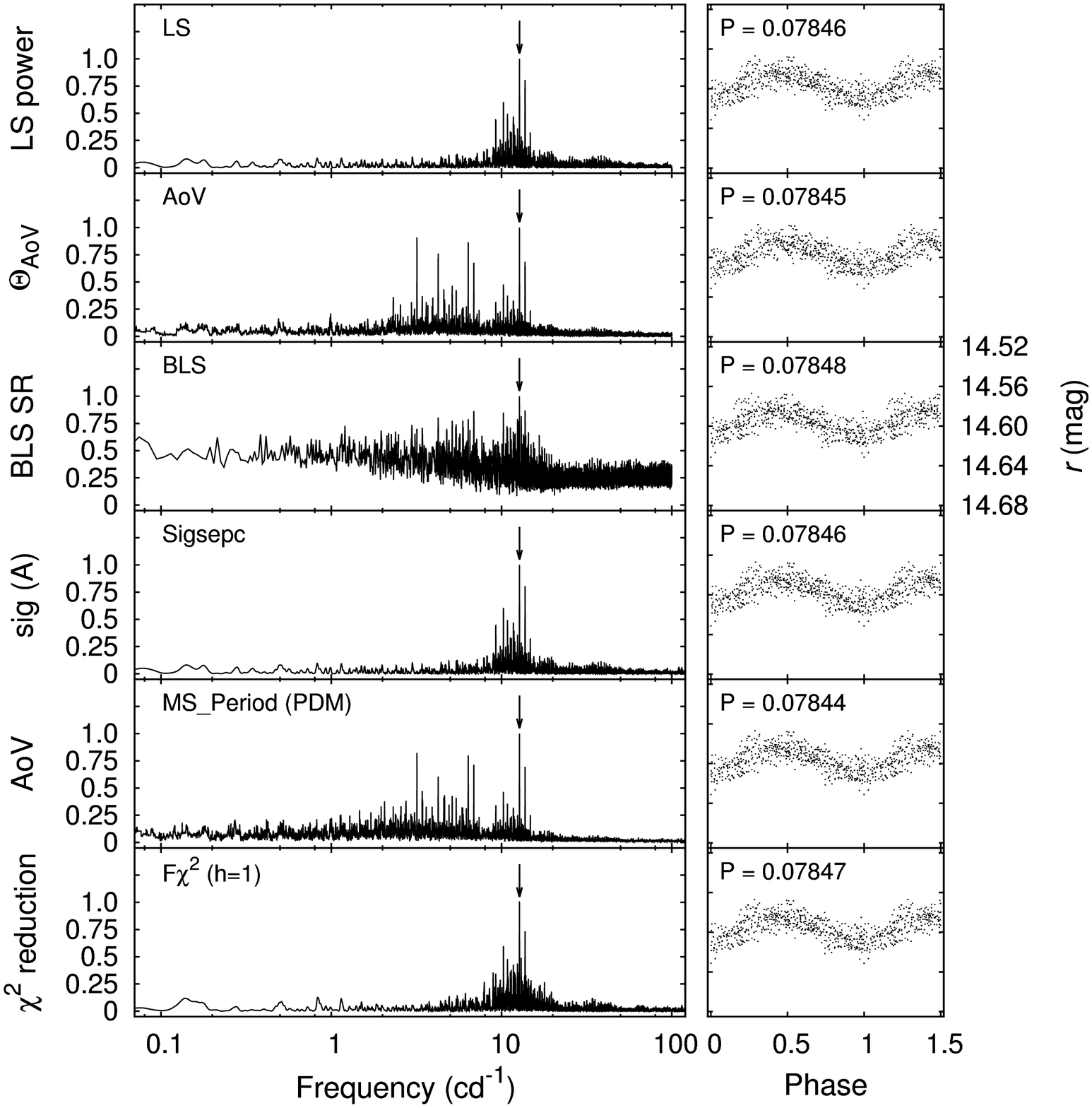}}  
  \subfloat[]{\includegraphics[width=0.34\linewidth, angle=0]{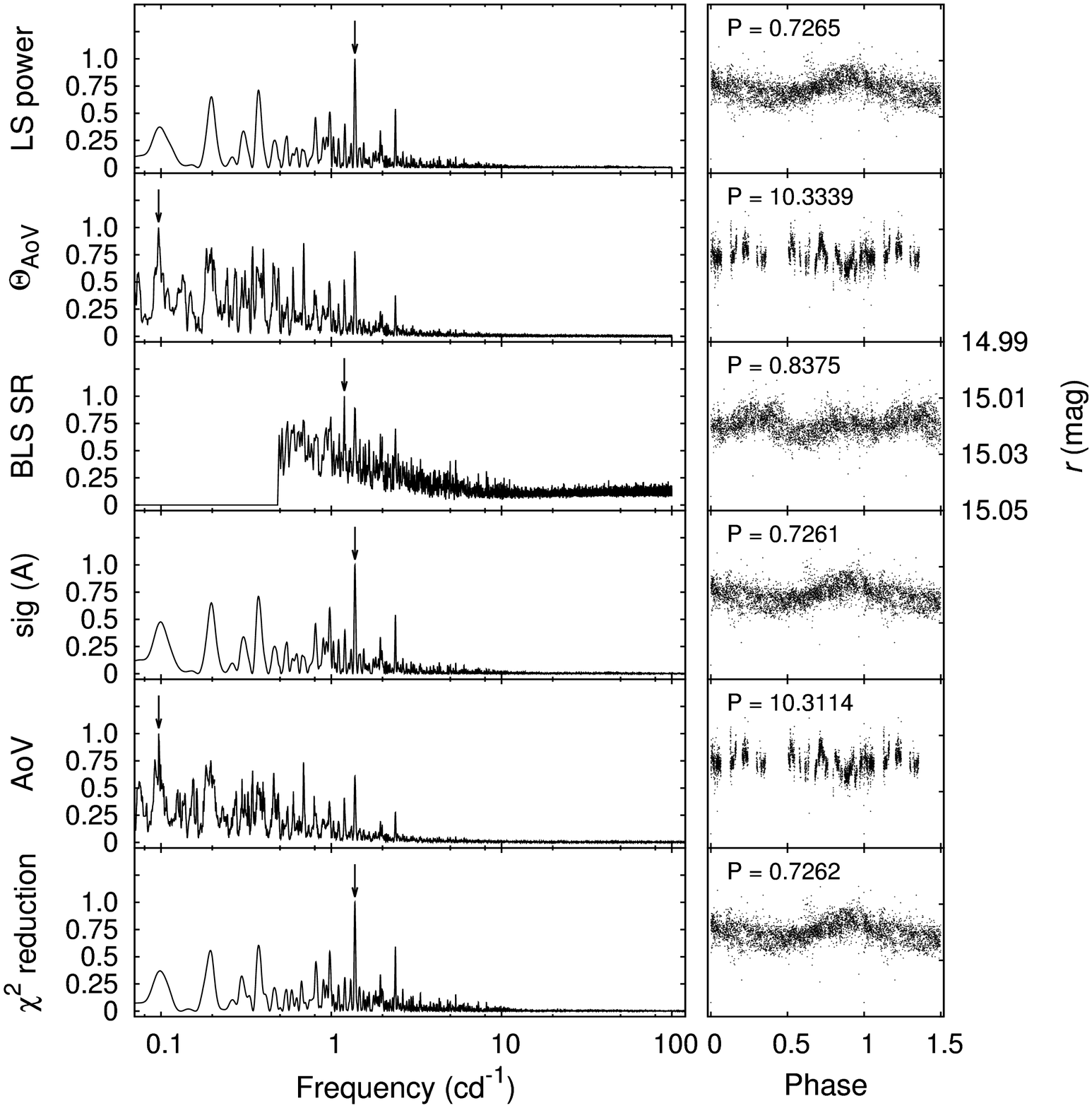}}
  \subfloat[]{\includegraphics[width=0.34\linewidth, angle=0]{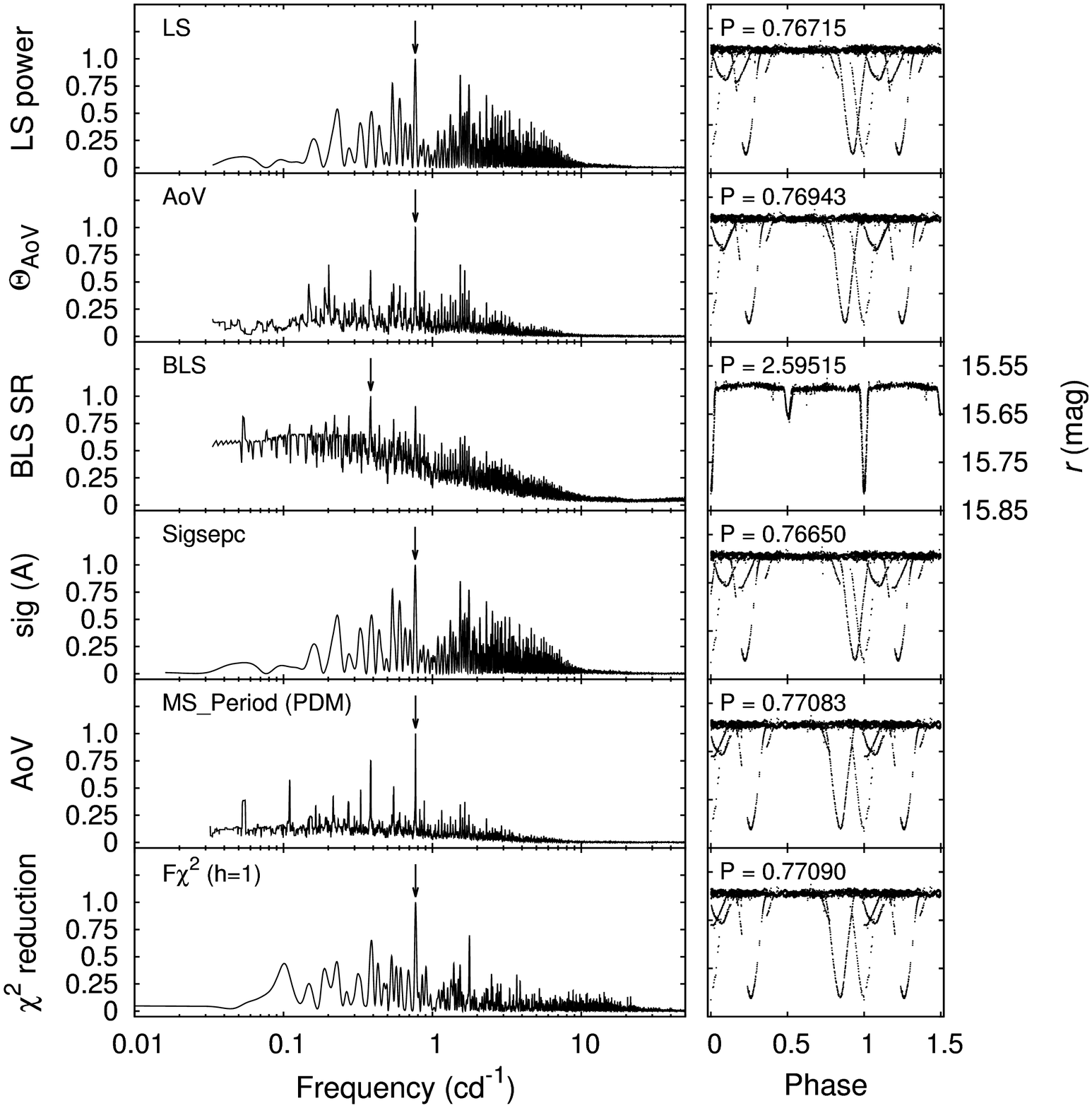}}\vspace*{-0.8em}
  \caption{Examples of power spectrum measurement and its corresponding phased diagram with period at the highest peak (arrows) in power spectrum.  The scales of $y$-axis are normalized to the maximum value.  One case shows that all periods are identical within errors (left panel; V2256), while in the other cases the derived period differs a lot.}
  \label{Fig4}
\end{figure*}

% Section 3.3
\subsection{Periodicity analysis}
For periodicity, we first used the {\sc VARTOOLS} light curve analysis program to determine likely periods of variable stars\footnote{\url{http://www.astro.princeton.edu/~jhartman/vartools.html}}.  This code computes three different periodograms for the light curves;(i) Lomb-Scargle (LS: \citealt{lom76,sca82,pre89,pre92}), (ii) Analysis of Variance (AoV: \citealt{sch89,dev05}), and (iii) Box-fitting Least-Squares (BLS: \citealt{kov02}) algorithms.  Because each method is sensitive to different types of periodic variables, it is often useful to combine results from many different analysis of periodicity.  Basically we adopt the similar treatment as in \citet{har08b}.  We limited the search range of these algorithms to periods between 0.01 and 30 days with relatively fine frequency resolution of less than 0.01/$T$ ($T$ is the time-span of the light curve).

We also obtained period estimates from the statistical technique {\sc SigSpec} \citep{ree07}\footnote{\url{http://homepage.univie.ac.at/peter.reegen/download.html}}.   It calculates the spectral significance (sig) levels based on the analytical solution for discrete Fourier transform amplitude spectrum, including dependencies on frequency and phase.  In recent years, this package is mainly used in the study of stellar pulsation (i.e., asteroseismology), which shows up as small, periodic variations in the brightness of the stars (e.g., \citealt{gru07,ben10,gug12}).  In order to further crosscheck results of periodicity analysis, we also performed a multi-step period search algorithm {\sc MS\textunderscore Period} \citep{shi04}\footnote{\url{http://www.physics.ox.ac.uk/users/msshin/science/code/MultiStep_Period/}}.  

Figure \ref{Fig4} shows examples of three selected light curves; power spectrum and phased diagram corresponding to the highest peak.  The differences in the various algorithms depend strongly on the shape of variation.  One case (V2256) shows that all periods are essentially identical ($P\simeq$ 0.0784 days), while in the other cases the derived period differ a lot.  Visual inspections are necessary for V2173 and V1482 to choose the most plausible periods; $P\sim0.726$ and 2.595 respectively.
    
% Section 4
\section{New catalog of variable stars in M37 field}
We provide a catalog of the total 2306 stars that exhibits convincing variations that are induced by flares, pulsations, eclipses, starspots, or in some cases, unknown causes.  This catalog is made available in electronic form at \url{http://stardb.yonsei.ac.kr/}.  Table \ref{TabA1} gives the basic information about previously known and new variable stars that we detected in the field of M37.  The columns are as follows: 

{\underline{\it VarID}}: The variable star IDs for the first 1483 stars are from the numbering system of \citet{har08b}.  New variable ID numbers are given in ascending order by their right ascension from V1484 to V2306.

{\underline{\it Object designation}}: Each object is designated by the combination of right ascension and declination for J2000 epoch (e.g., 055536.78+323345.97).

{\underline{\it Period}}: We basically used the dominant periods which were highly scored by several methods.  For EBs, we checked the existence of secondary eclipse when phased at twice the period.  In cases where our periodicity analysis is affected by a large scatter of light curves due to photometric bias or source blending, we used periods taken from the previous catalog \citep{har08b}, and then denoted by $\dag$ in that column.  We also found evidences of multiperiodicity in some pulsating stars (see Section 5.2).
  
{\underline{\it Amplitude}}: The range of variability for periodic variables was derived from phase diagram in which the shape of the light curve was modeled by equation (2).  This amplitude is defined as the full-amplitude of model function ($\mathrm{Amp}$ = $m_{max} - m_{min}$).  When this was not possible, i.e., cases of EBs and irregulars, we used the original light curve to estimate the amplitude.
  
{\underline{\it Magnitudes}}: $ugrizJHK_s$ magnitudes for many of our stars are available from the Sloan Digital Sky Survey Data Release 7 ({\sc SDSS} DR7)\footnote{\url{http://www.sdss.org/dr7/}} and Two Micro All Sky Survey ({\sc 2MASS})\footnote{\url{http://www.ipac.caltech.edu/2mass/}}.  Each catalog is positionally matched to {\sc MMT} sources with an average distance of 0.6$^{\prime\prime}$.  In the case of {\sc SDSS}, the photometric quality at low Galactic latitudes is not guaranteed to be accurate to the {\sc SDSS} quoted limits of 2\% in color, and photometric depths are about two magnitude shallower than our data.  Similarly, {\sc 2MASS} is not deep enough to detect low-mass members in the cluster.

{\underline{\it VarType}}: We assigned five variability types similar to the variability classification scheme proposed by the GCVS \citep{sam09}.  Variable stars are grouped according to major astrophysical reasons for variability, such as flare (F), pulsating (P), rotating (R), eclipsing binary (EB), and aperiodic variables (A), including variable candidates (:).  Further details are given in Section 5.  If a variable belongs to more than one type of variability, it is indicated by combination of individual types, e.g., F+EB, P+EB:.

\begin{figure}[!t]
  \centering
  \includegraphics[width=1.0\linewidth, angle=0]{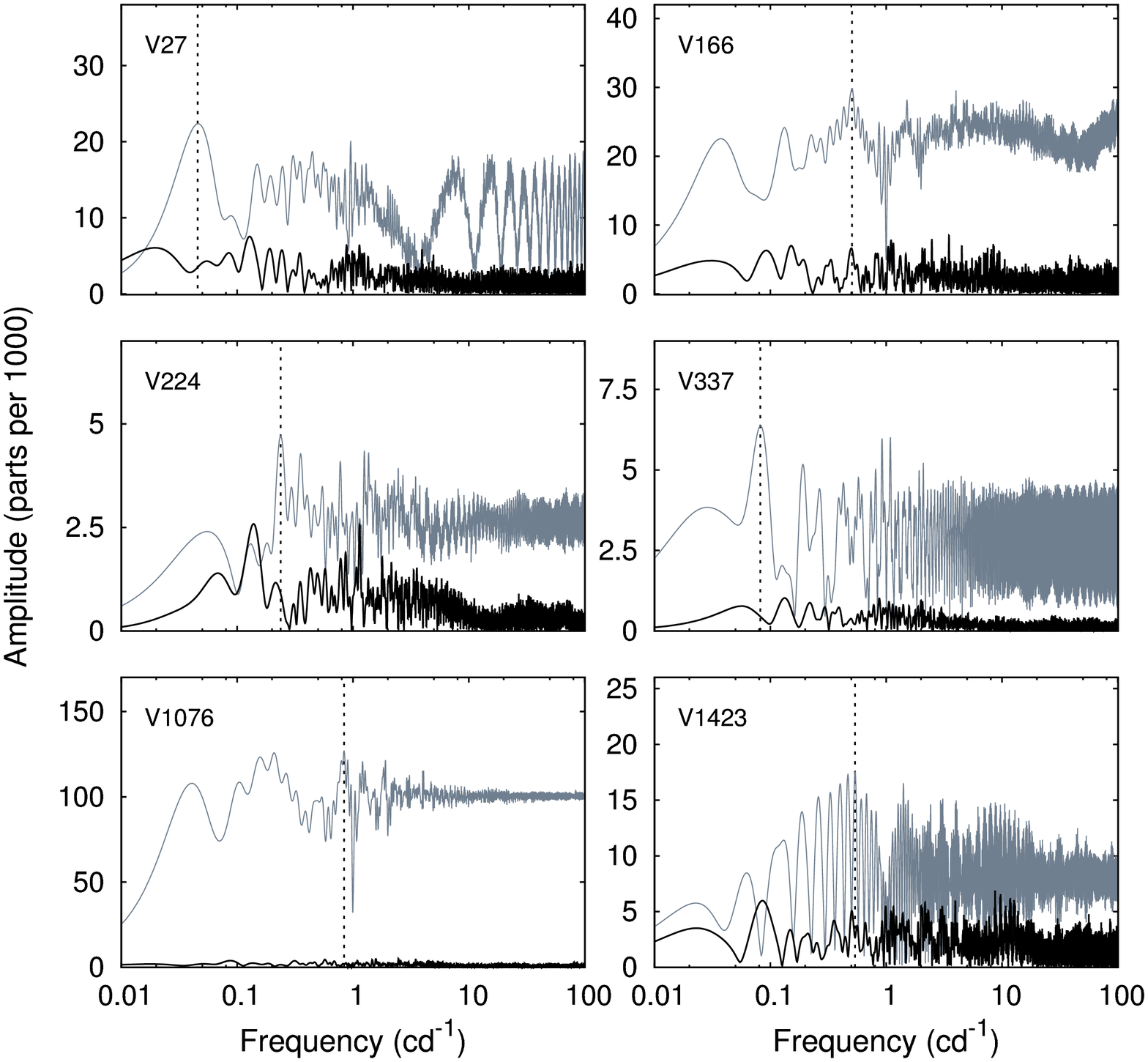}
  \caption{Amplitude spectrum of the archival (gray) and our new light curves (black).  Mean noise levels are greatly reduced in new analysis.}
  \label{Fig5}
\end{figure}

A total of 30 stars listed as variables in \citet{har08b} turned out to be not real variable in our study.  Their new amplitude spectrum often does not show anything significant (Figure \ref{Fig5}) and no apparent periodicity in the light curve.  These false positives are listed in Table \ref{Tab1}.

% Table 1
\begin{deluxetable}{rcrcrr}
\tabletypesize{\scriptsize}
\tablecaption{List of 30 stars whose variability turn out to be false\label{Tab1}}
\tablewidth{0pt}
\tablehead{
\colhead{} & \colhead{} & \colhead{StarID} & \colhead{Chip} & \colhead{$r$} & \colhead{Period\tablenotemark{a}}\\
\colhead{No.} & \colhead{VarID} & \colhead{(J2000)} & \colhead{(\#)} & \colhead{(mag)} & \colhead{(days)}
}
\startdata
 1\phn & V27 \phn &   055118.34+323201.71 \phn & 32\phn & 20.938\phn & 16.364\phn \\
 2\phn & V85 \phn &   055125.32+324341.29 \phn & 28\phn & 20.862\phn & 7.668\phn \\
 3\phn & V88 \phn &   055125.46+324315.32 \phn & 28\phn & 21.820\phn & 1.646\phn \\
 4\phn & V120 \phn &   055128.52+322339.10 \phn & 35\phn & 17.196\phn & 6.859\phn \\
 5\phn & V155 \phn &   055131.10+323756.98 \phn & 30\phn & 16.847\phn & 7.424\phn \\
 6\phn & V166 \phn &   055131.96+323600.24 \phn & 31\phn & 20.895\phn & 10.451\phn \\
 7\phn & V224 \phn &   055137.29+324011.78 \phn & 29\phn & 18.165\phn & 4.571\phn \\
 8\phn & V230 \phn &   055138.02+324011.65 \phn & 29\phn & 16.690\phn & 4.595\phn \\
 9\phn & V250 \phn &   055139.59+324013.27 \phn & 29\phn & 17.282\phn & 4.551\phn \\
10\phn & V330 \phn &   055151.25+323527.70 \phn & 22\phn & 16.831\phn & 5.428\phn \\
11\phn & V337 \phn &   055151.72+323510.02 \phn & 22\phn & 17.920\phn & 5.442\phn \\
12\phn & V342 \phn &   055152.18+323522.44 \phn & 22\phn & 19.061\phn & 5.438\phn \\
13\phn & V347 \phn &   055152.55+323536.03 \phn & 22\phn & 15.574\phn & 5.500\phn \\
14\phn & V367 \phn &   055153.88+323459.77 \phn & 22\phn & 17.463\phn & 5.515\phn \\
15\phn & V390 \phn &   055155.42+323523.56 \phn & 22\phn & 18.268\phn & 5.481\phn \\
16\phn & V408 \phn &   055156.77+323534.61 \phn & 22\phn & 15.733\phn & 5.481\phn \\
17\phn & V433 \phn &   055157.99+323528.59 \phn & 22\phn & 16.617\phn & 5.481\phn \\
18\phn & V492 \phn &   055202.16+324517.30 \phn & 19\phn & 20.849\phn & 16.586\phn \\
19\phn & V973 \phn &   055235.48+322104.58 \phn & 18\phn & 16.847\phn & 5.089\phn \\
20\phn & V1076 \phn &   055241.54+322107.71 \phn & 18\phn & 20.262\phn & 10.927\phn \\
21\phn & V1237 \phn &   055258.14+322152.09 \phn & 9\phn & 16.775\phn & 5.692\phn \\
22\phn & V1260 \phn &   055300.53+323445.47 \phn & 4\phn & 19.851\phn & 15.895\phn \\
23\phn & V1282 \phn &   055303.05+323621.95 \phn & 4\phn & 21.530\phn & 1.666\phn \\
24\phn & V1305 \phn &   055305.05+323447.83 \phn & 4\phn & 18.166\phn & 7.966\phn \\
25\phn & V1330 \phn &   055306.74+323442.52 \phn & 4\phn & 18.061\phn & 8.143\phn \\
26\phn & V1371 \phn &   055310.61+323748.13 \phn & 3\phn & 20.902\phn & 9.934\phn \\
27\phn & V1372 \phn &   055310.65+323740.82 \phn & 3\phn & 20.985\phn & 10.633\phn \\
28\phn & V1400 \phn &   055312.89+323005.46 \phn & 6\phn & 16.770\phn & \nodata\phn \\
29\phn & V1421 \phn &   055314.64+322656.12 \phn & 7\phn & 20.735\phn & 15.013\phn \\
30\phn & V1423 \phn &   055315.29+323051.00 \phn & 6\phn & 20.728\phn & 12.388\phn \\
\enddata
\tablenotetext{a} {Period is obtained by \citet{har08b}.}
\end{deluxetable}

% Figure 5

% Section 4.1
\subsection{Variability fractions}
In this work, we found 823 new variables (V1484--V2306) and identified types of their variability.   The discovery rate of new variables is increased by 60\% in comparison with the existing catalog.  The new variables are primarily 436 flaring stars, 30 sharp EBs, 48 low-amplitude pulsating variables, and 65 aperiodic variables, as well as many other rotating variables.

\begin{figure}[!t]
\centering
  \includegraphics[width=0.45\textwidth, angle=0]{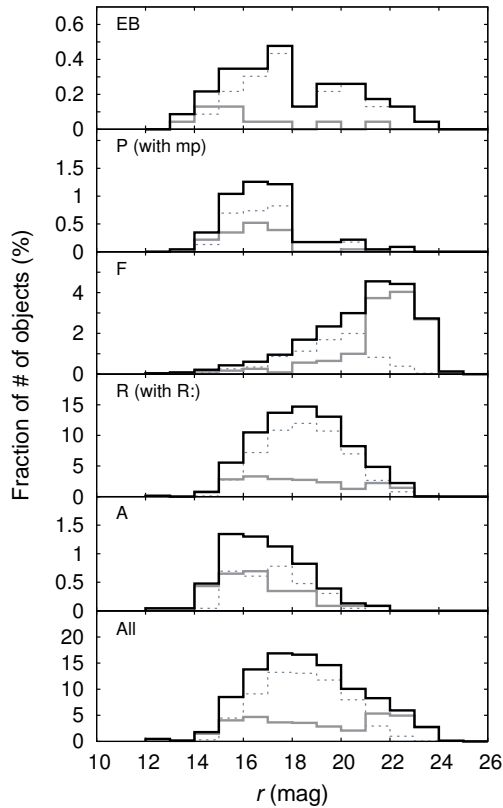}
  \caption{Variability fraction as a function of $r$ magnitude.  Previously known variables (V1--V1483), newly detected variables (V1484--V2306), and total number of observed stars are marked with dashed lines, gray solid lines, and black solid lines, respectively.}
  \label{Fig6}
\end{figure}

\begin{figure}[!t]
\centering
  \includegraphics[width=0.5\textwidth, angle=0]{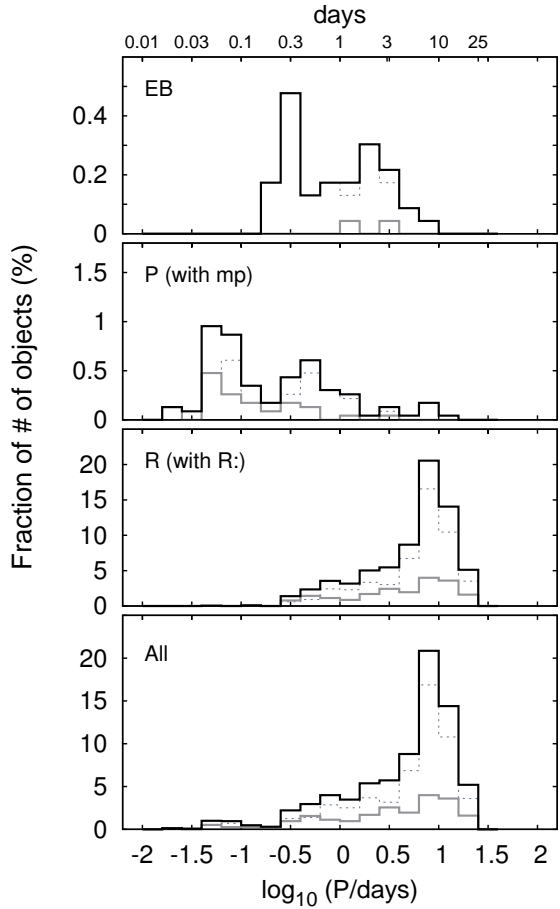}
  \caption{Variability fraction as a function of period.  The representation of the lines is same as Figure \ref{Fig6}.}
  \label{Fig7}
\end{figure}

We briefly explored how variability fractions depends on their brightness/period and variable types.  As shown in Figure \ref{Fig6}, short-period, low-amplitude pulsating stars and episodic flare events are dominantly found at the bright- and faint-end of magnitude, respectively.  In Figure \ref{Fig7}, we plot variability fraction as a function of period.  The period range of new variables is very similar to that reported by \citet{har08b}.  Of these, most of new pulsating candidates show $\delta$ Scuti-like characteristics ($P < $ 0.3 days).  Also the majority of periodic variability can be explained by rotational modulation of surface features (i.e., starspots).

% Figure 6 

% Figure 7

% Figure 8
\begin{figure}[!t]
  \subfloat[]{\includegraphics[width=\linewidth, angle=0]{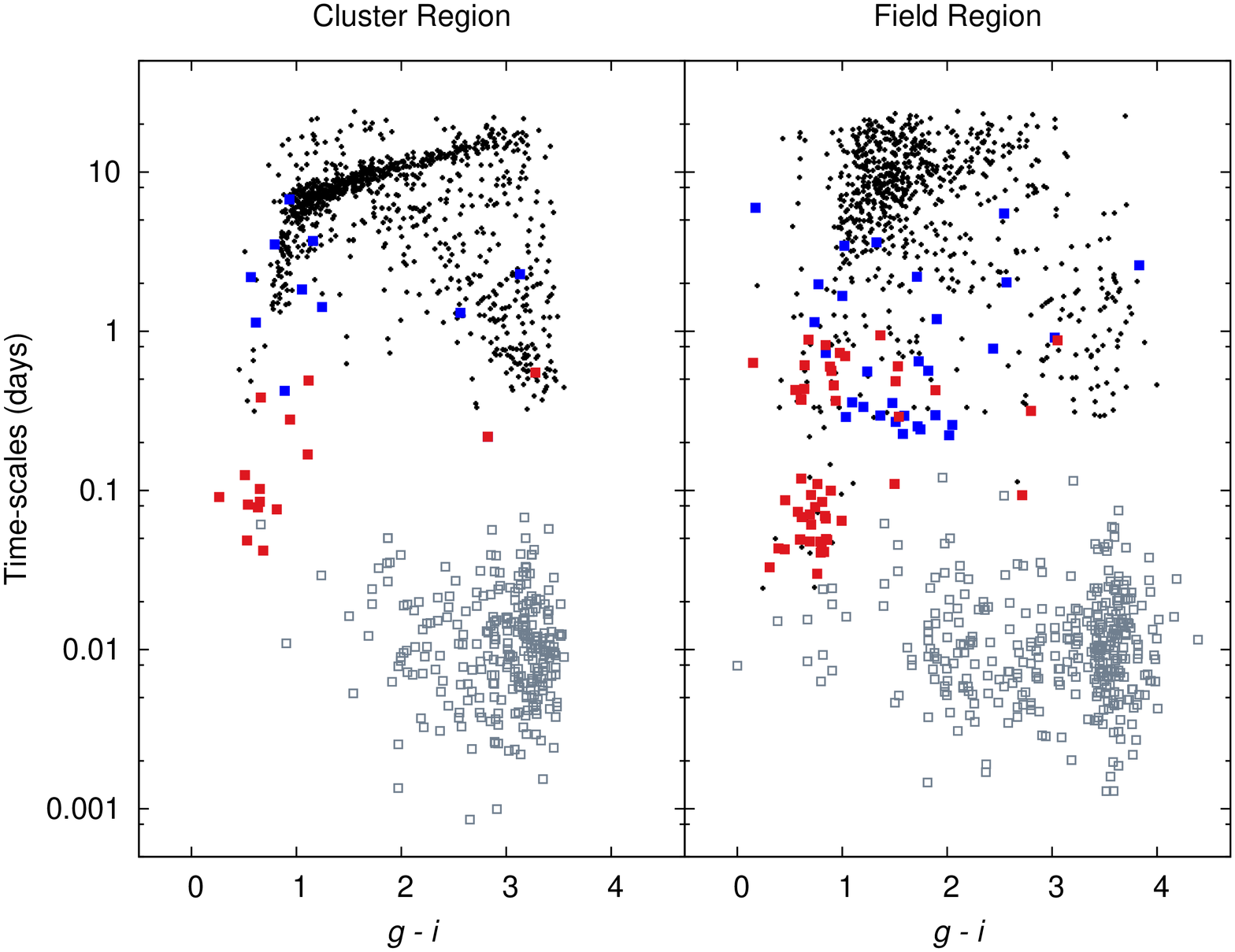}}\vspace*{-1.3em}
  \subfloat[]{\includegraphics[width=\linewidth, angle=0]{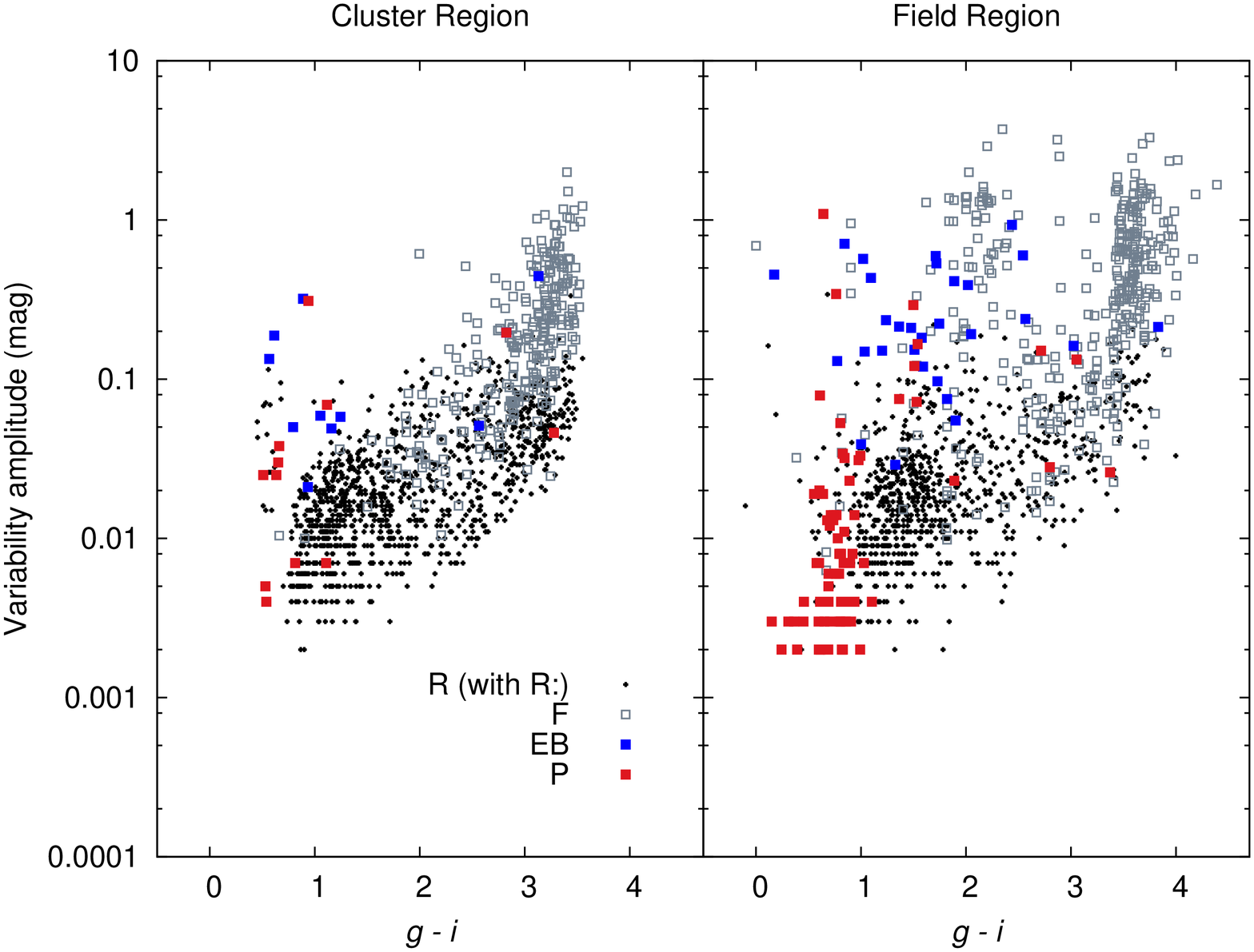}}\vspace*{-0.7em}
  \caption{$g - i$ color vs. variability time-scales and amplitude relations for all periodic variables (black points) in the cluster and field region, respectively.  For comparison purposes, we also show the parameter distributions of 625 flare events (gray squares).  The flare time-scale is defined as the time duration of flare events at 80\% of its peak level ($\tau_{0.2}$), and the flare amplitude is magnitude at the moment of flare maximum.}
  \label{Fig8}
\end{figure}

% Section 4.2
\subsection{Time-scales and amplitudes of the variability}
Time-scales and amplitudes of variability are intrinsic properties of the star and are independent of the distance \citep{eye08}.  These two quantities can easily be measured from the light curves.  We extracted the basic properties (e.g., period, amplitude) of all periodic variables using the Fourier decomposition technique.  For comparison purposes, we utilized the flare parameters: (i) flare duration $\tau_{0.2} (=t_{0.2,rise} + t_{0.2,decay})$ that is defined as the time-scales of flare events at 80\% of its peak level, and (ii) flare amplitude $\Delta$m$_{peak}$ that is magnitude at the moment of flare maximum (S. -W. Chang et al., in preparation, hereafter Paper III).

We separate the variable objects into photometrically selected candidate cluster and field members provided by \citet{har08b}, respectively.  Since the $g - i$ color covers a relatively wide range from $-0.5$ to 4.5, we used $g - i$ as a separator between a sequence of cluster and that of field population.   As can be seen in Figure \ref{Fig8}, short period ($<$ 1 day), small amplitude pulsating stars are found at the blue-end of the distribution of both cluster and field sources, while episodic flare events are dominantly found from the sources in the red color range.

Our new data strengthen the relationship between the color of stars and their variability parameters derived for cluster members previously (Figure \ref{Fig8}).  The cluster members show a tight and nearly linear relation between color index and rotation period in the range of 1 $<$ $\log P$ $<$ 20 days, although the change in slope occurs after the knee point at $P$ $\simeq$ 7 days.  This sequence forms a curved, diagonal band from bright blue stars to faint red ones, whose periods are increasing with increasing color index (or magnitude), while for non-cluster members there is no clear correlation.
  
The color-amplitude relation for the cluster members shows that, statistically, cooler stars have greater amplitudes in the case of both periodic variables and flares.  As in the previous study on stellar rotation (see Section 6 and Figures 14, 15 and 17 in \citealt{har09a}), there is a strong correlation between Rossby number and amplitude, which can be interpreted as stars rotating faster having more spots, larger spots, and/or greater spot-photosphere contrast than slower rotating stars.  Cooler stars have on average lower Rossby numbers at a particular age, so they end up having larger amplitudes.

% Section 4.3
\subsection{Color-magnitude diagram of variable stars}
\begin{figure}[!t]
\centering
  \includegraphics[width=\linewidth, angle=0]{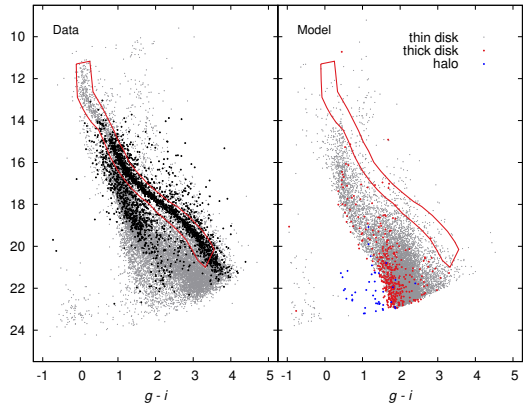}  
  \caption{Color-magnitude diagram with marked positions of variables (black dots) found in the M37 field.  We compare the $g - i$ vs. $i$ diagram of both observed (left) and simulated (right) data.  Photometrically selected cluster region is plotted as red line in both panels.}
  \label{Fig9}
\end{figure}

Figure \ref{Fig9} illustrates the color-magnitude diagram (CMD) for all objects and all variable stars in the M37 field.  In the new catalog, we found that about half of variable objects are distributed along the cluster main-sequence, and the other half are located in the field or in the outskirts of the cluster sequence.  In order to identify the latter stars, we compared our data with the {\sc TRILEGAL}\footnote{The latest version of TRIdimensional modeL of thE GALaxy is available at \url{http://stev.oapd.inaf.it/cgi-bin/trilegal}.} model to reproduce the CMD at this location of the Galaxy.  Most of the input parameters are described in the \citet{gir05}.  For this analysis, we set the pointing parameters for the field center ($l=177.635$; $b = 3.091$) and total field area ($\square$ = 0.16 deg$^{2}$), a limiting magnitude as $r$ = 23.5 and a Salpeter initial mass function with cutoff at 0.01 M$_{\sun}$.  The remaining parameter values are fixed to default values that correspond to their model calibration.  The model output consists mostly of thin disk stars (96\%) with small contribution of old stars in the thick disk and halo (4\%).  Non-member, variable stars may be related in some way to the thin disk population, but their distributions are not fully explained by the model.

% Figure 9

% Section 5
\section{Types of variability}
In the following subsections, we investigate the nature of each variability class: (i) 61 EB systems, (ii) 92 multiperiodic variable stars, (iii) 132 aperiodic variables, and (iv) 436 flare stars, respectively.  In order to understand variability properties, we utilized both color-magnitude and color-color diagrams to separate variable objects into broad magnitude ranges or spectral classes, to the extent possible.

% Section 5.1
\subsection{Eclipsing binary star systems}
\begin{figure*}[!t]
\begin{center}
  \includegraphics[width=\linewidth, angle=0]{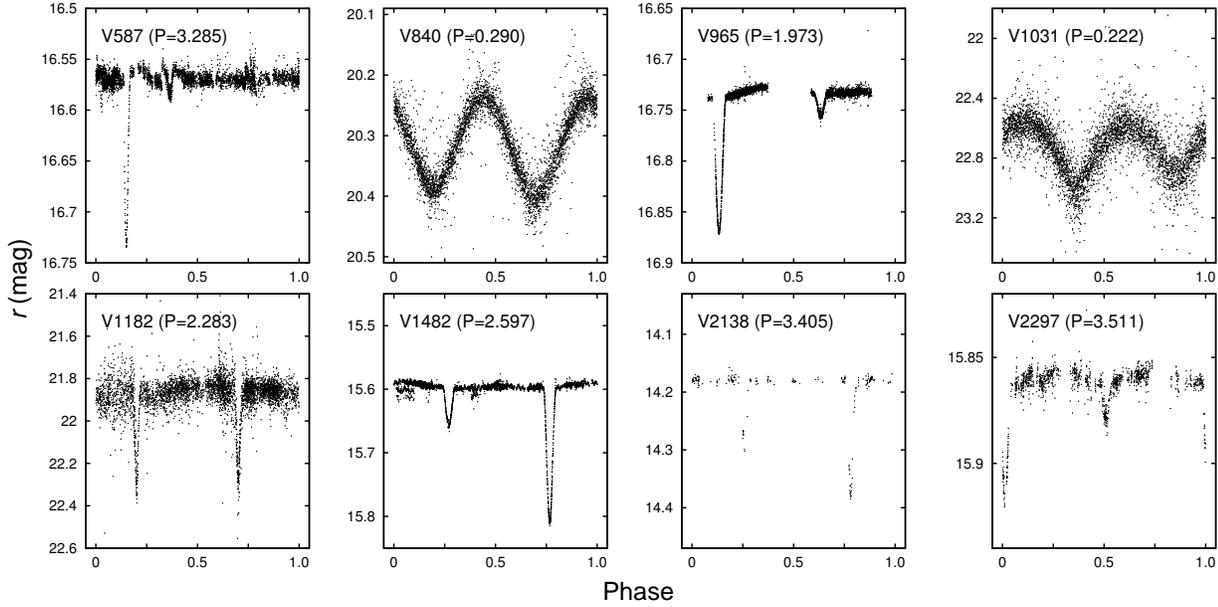}  
  \caption{Phased light curves of 8 newly discovered eclipsing binaries.}
  \label{Fig10}
\end{center}
\end{figure*}

Our photometric time-series data provide the light curves of 61 EBs.  Figure \ref{Fig10} shows phased light curves of 8 newly discovered EBs.  We categorize EBs into three groups according to the shape of phased light curves, and discuss each group below.

% Figure 10

% Section 5.1.1
\subsubsection{Detached or Semi-detached EBs}
EBs in this category are either Algol-type systems or $\beta$ Lyrae-type systems.  To derive their geometric parameters, we used the Detached Eclipsing Binary Light curve (DEBiL) fitter by \citet{dev05}\footnote{\url{https://www.cfa.harvard.edu/~jdevor/DEBiL.html}}, which employs a simple model of a perfectly detached binary system in a classical 2-body orbit.  Table \ref{Tab2} lists parameters derived for all detached/semi-detached binaries in the M37 field.  EBs with well determined periods are included in DEBiL analysis.  Special notes are given below for interesting features.

\paragraph{V2} This bright EB system was missed by previous study due to magnitude cut, but it clearly shows evidences of eclipse in our new light curve.

\paragraph{V587} The light curve of this EB system contains two distinct periodic signals, the 3.6173 days modulation with a small amplitude ($\Delta m$ $\simeq$ 0.056 mag) and an additional period at 3.2857 days (orbital period).  Since an ellipsoidal variation should have the same period as the orbit, the first signal is more likely a consequence of starspots and rotational modulation..  After 3.6173-d modulation is removed, the light curve clearly shows the signature of a typical Algol-type binary with both primary and secondary eclipses.

\paragraph{V1482} This EB system is a newly recognized Algol-type binary with a $\delta$ Sct-type pulsating component.  We have clearly detected an orbital period of 2.5972 days including multiple primary and secondary eclipses.  The short period variability caused by stellar pulsation is shown in the out-of-eclipse light curves.  In order to determine the oscillation modes of the pulsation, a frequency analysis was performed in the out-of-eclipse data using the program {\sc PERIOD04} \citep{len05}.  Our analysis not only shows peaks as a consequence of binary effects ($f_{1}$ = 2.5638 and $f_{2}$ = 1.3006) but also reveals the presence of the dominant pulsation frequency $f_{3}$ = 0.0465 with a small amplitude ($\Delta m$ = 0.0035 mag).

% Table 2
\begin{deluxetable}{rcrrrrrrrrrrl}
\tabletypesize{\scriptsize} %\tiny}
\rotate
\tablecaption{Parameters for 22 Detached and Semi-detached EB systems\label{Tab2}}
\tablewidth{0pt}
\tablehead{
\colhead{   } & \colhead{     } & \colhead{$r$} & \colhead{$P_{orb}$} & \colhead{} & \colhead{    } & \colhead{    } & \colhead{$m_{1}$} & \colhead{$m_{2}$} & \colhead{} & \colhead{} & \colhead{$\omega$} & \colhead{} \\ 
\colhead{No.} & \colhead{VarID} & \colhead{(mag)} & \colhead{(days)} & \colhead{$e$\tablenotemark{a}} & \colhead{$R_{1}/a$} & \colhead{$R_{2}/a$} & \colhead{(mag)} & \colhead{(mag)} & \colhead{sin($i$)} & \colhead{$t_{0}$\tablenotemark{b}} & \colhead{(deg)} & \colhead{Note\tablenotemark{c}}
}
\startdata
1\phn & \object[V* V541 Aur]{V1}\phn & 13.677\phn & \nodata\phn & \nodata & \nodata & \nodata & \nodata & \nodata & \nodata & \nodata & \nodata & Not enough data \phn\\
2\phn & \object[V* V540 Aur]{V2}\phn & 14.840\phn & \nodata\phn & \nodata & \nodata & \nodata & \nodata & \nodata & \nodata & \nodata & \nodata & Not enough data \phn\\
3\phn & \object[Cl* NGC 2099    HGH V55]{V55}\phn & 22.549\phn & 0.7780\phn & $\le$0.08\phn & 0.17\phn & 0.15\phn & 22.81\phn & 24.12\phn & 1.00\phn & 0.65\phn & 283.4\phn & D\phn\\
4\phn & \object[Cl* NGC 2099    HGH  330224]{V146}\phn & 20.463\phn & 0.6477\phn & \nodata & \nodata & \nodata & \nodata & \nodata & \nodata & \nodata & \nodata & SD; No secondary minimum? \phn\\
5\phn & \object[NGC 2099    HGH V226]{V226}\phn & 22.121\phn & 0.9123\phn & $\le$0.01\phn & 0.10\phn & 0.04\phn & 22.23\phn & 24.59\phn & 1.00\phn & 0.53\phn & 129.5\phn & D \phn\\
6\phn & \object[NGC 2099    HGH V395]{V395}\phn & 17.392\phn & 3.4460\phn & 0.01\phn & 0.10\phn & 0.08\phn & 17.96\phn & 18.35\phn & 1.00\phn & 0.33\phn & 232.1\phn & D\phn\\
7\phn & \object[NGC 2099    HGH V457]{V457}\phn & 19.474\phn & 2.1995\phn & $\le$0.08\phn & 0.08\phn & 0.07\phn & 20.15\phn & 20.30\phn & 1.00\phn & 0.75\phn & 239.7\phn & D\phn\\
8\phn & \object[2MASS J05520131+3232509]{V485}\phn & 17.080\phn & 1.6676\phn & \nodata & \nodata & \nodata & \nodata & \nodata & \nodata & \nodata & \nodata & D; No secondary minimum? \phn\\ 
9\phn & \object[Cl* NGC 2099    HGH V587]{V587}\phn & 16.572\phn & 3.2857\phn & 0.52\phn & 0.06\phn & 0.02\phn & 16.58\phn & 21.24\phn & 1.00\phn & 0.31\phn & 209.4\phn & D + RM (P=3.6173 days) (new)\phn\\
10\phn & \object[Cl* NGC 2099    HGH V692]{V692}\phn & 16.669\phn & 1.8324\phn & 0.06\phn & 0.28\phn & 0.06\phn & 16.73\phn & 19.80\phn & 0.98\phn & 0.01\phn & 77.2\phn & SD \phn\\
11\phn & \object[Cl* NGC 2099    HGH V716]{V716}\phn & 21.635\phn & 2.0272\phn & \nodata & \nodata & \nodata & \nodata & \nodata & \nodata & \nodata & \nodata & D; No secondary minimum?\phn\\
12\phn & \object[Cl* NGC 2099    HGH V733]{V733}\phn & 17.383\phn & 0.7310\phn & $\le$0.04\phn & 0.43\phn & 0.34\phn & 17.61\phn & 18.91\phn & 0.99\phn & 0.77\phn & 73.3\phn & D \phn\\ % Period and amplitude variations
13\phn & \object[Cl* NGC 2099    HGH V965]{V965}\phn & 16.736\phn & 1.9732\phn & 0.00\phn & 0.16\phn & 0.05\phn & 16.75\phn & 21.05\phn & 1.00\phn & 0.27\phn & 357.0\phn & SD (new) \phn\\
14\phn & \object[2MASS J05523843+3223296]{V1028}\phn & 18.730\phn & 5.4956\phn & 0.03\phn & 0.04\phn & 0.04\phn & 19.21\phn & 19.84\phn & 1.00\phn & 0.55\phn & 267.5\phn & D + RM\phn\\
15\phn & \object[2MASS J05524614+3239336]{V1141}\phn & 16.977\phn & 3.6948\phn & $\le$0.66\phn & 0.06\phn & 0.01\phn & 16.98\phn & 22.60\phn & 1.00\phn & 0.29\phn & 268.3\phn & D\phn\\
16\phn & \object[Cl* NGC 2099    HGH V1182]{V1182}\phn & 21.869\phn & 2.2834\phn & 0.05\phn & 0.07\phn & 0.06\phn & 22.40\phn & 22.87\phn & 1.00\phn & 0.16\phn & 268.3\phn & D (new)\phn\\
17\phn & \object[Cl* NGC 2099    HGH V1187]{V1187}\phn & 19.742\phn & 1.1919\phn & $\le$0.41\phn & 0.12\phn & 0.02\phn & 19.77\phn & 23.56\phn & 1.00\phn & 0.86\phn & 271.6\phn & D\phn\\
18\phn & \object[2MASS J05531108+3224434]{V1380}\phn & 14.874\phn & 2.1898\phn & $\le$0.05\phn & 0.16\phn & 0.06\phn & 15.01\phn & 17.19\phn & 0.99\phn & 0.84\phn & 280.1\phn & D + $\gamma$ Doradus component (P=0.941/1.174 days)\phn\\
19\phn & \object[Cl* NGC 2099    HGH S2]{V1456}\phn & 17.006\phn & 5.9680\phn & 0.12\phn & 0.07\phn & 0.07\phn & 17.47\phn & 18.15\phn & 1.00\phn & 0.54\phn & 89.3\phn & D\phn\\
20\phn & V1482\phn & 15.597\phn & 2.5972\phn & 0.02\phn & 0.14\phn & 0.08\phn & 15.71\phn & 18.14\phn & 0.99\phn & 0.17\phn & 76.1\phn & D + $\delta$ Scuti component (P=0.0465 days) (new)\phn\\
21\phn & V2138\phn & 14.181\phn & 3.4051\phn & 0.04\phn & 0.09\phn & 0.04\phn & 14.28\phn & 16.85\phn & 1.00\phn & 0.63\phn & 191.7\phn & D (new)\phn\\
22\phn & V2297\phn & 15.863\phn & 3.5109\phn & $\le$0.01\phn & 0.12\phn & 0.02\phn & 15.88\phn & 20.42\phn & 1.00\phn & 0.32\phn & 229.7\phn & D (new)\phn\\
\enddata
\tablecomments{The column description is as follows. $e$: Orbital eccentricity; $R_{1}/a$: Radius of large star (in units of semimajor axis); $R_{2}/a$: Radius of small star (in units of semimajor axis); $m_{1}$: Brightness of large star (magnitudes); $m_{2}$: Brightness of small star (magnitudes); sin($i$): Sine of inclination; $t_{0}$: Phased epoch of periastron; $\omega$: Argument of periastron (in degrees).}
\tablenotetext{a}{For the cases where the errors are larger than $e$, we show the best-fit value plus the error as an upper limit.}
\tablenotetext{b}{Phased epoch of periastron: Modified Julian date folded by the period (see Table 4 in \citealt{dev05}).}
\tablenotetext{c}{D: detached eclipsing binary system; SD: semi-detached eclipsing binary system; RM: rotational modulation.}
\end{deluxetable}

% Section 5.1.2
\subsubsection{Contact EBs}
The light curves of contact binary stars (also known as W UMa-type variable stars; EW) are very much different from those for detached binaries.  Our new catalog lists 22 contact binaries of which four are new discoveries (V840, V1031, V1250, V1463).

\paragraph{V37, V706, V1031} These systems are one of the fastest orbiting contact binary systems ($P$ $\simeq$ 0.22 days).  It is known that observed distribution of orbital periods has a cut-off period limit at 0.22 days (e.g., \citealt{ruc92,ruc02,mol13}).  We have not found anything faster in our study.

% Figure 11
\begin{figure}[!t]
 \subfloat[]{\includegraphics[width=\linewidth, angle=0]{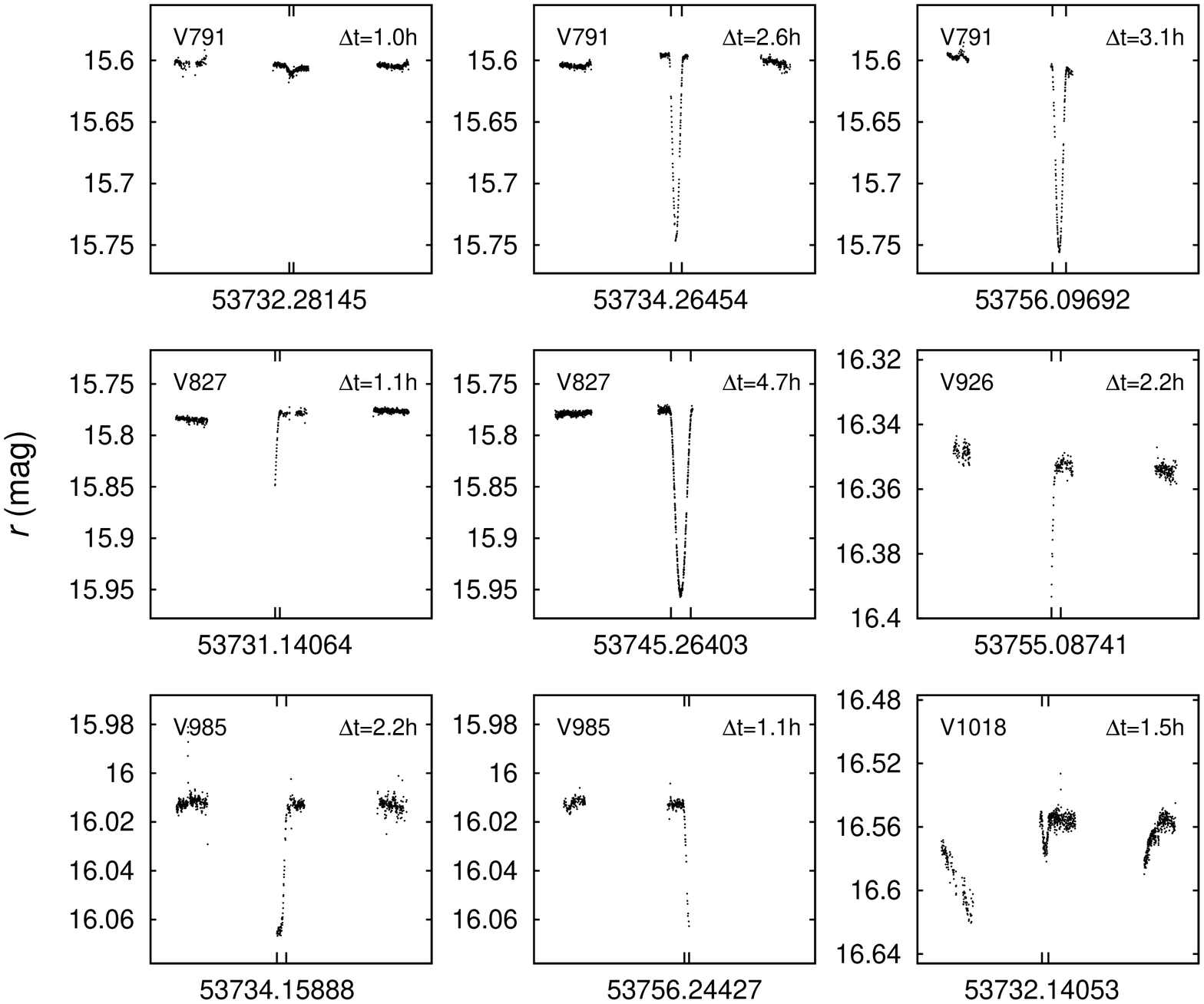}}\vspace*{-1.3em}
 \subfloat[]{\includegraphics[width=\linewidth, angle=0]{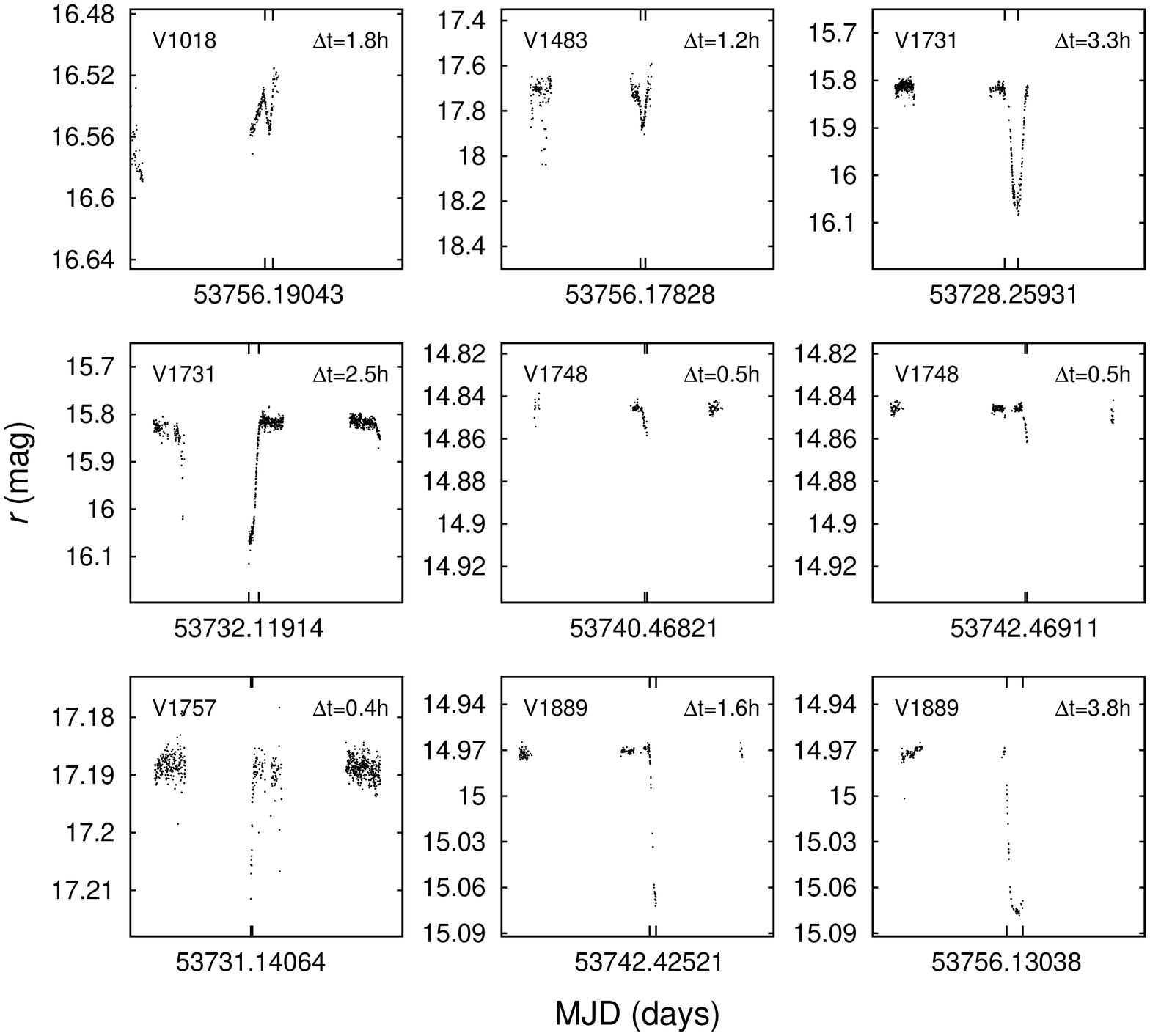}}\vspace*{-1.3em}
 \caption{Zoomed-in light curve collection of eclipse candidates.  Short vertical lines in the middle of each panel mark epochs of eclipses.  The numbers in the upper right denote the duration ($\Delta t$) between the start time of eclipse and the end.}
 \label{Fig11}
\end{figure}

% Table 3
\begin{deluxetable}{rccrcrrr}
\tabletypesize{\footnotesize}
\tablecaption{List of new 15 eclipsing binary candidates from CPA analysis\label{Tab3}}
\tablewidth{0pt}
\tablehead{
\colhead{} & \colhead{} & \colhead{Star} & \colhead{$r$} & \colhead{$N_\mathrm{event}$} & \colhead{$t_\mathrm{0}$\tablenotemark{a}} & \colhead{$\Delta t$\tablenotemark{b}} & \colhead{$d$\tablenotemark{c}} \\
\colhead{No.} & \colhead{VarID} & \colhead{(J2000)} & \colhead{(mag)} & \colhead{(\#)} & \colhead{(day)} & \colhead{(hour)} & \colhead{(mag)}
}
\startdata
1\phn & \object[2MASS J05522390+3228262]{V791}\phn & 055223.89+322826.47\phn & 15.604\phn & 1\phn & 53732.28145\phn & 1.0\phn & 0.011\phn \\
\phn & \phn & \phn & \phn & 2\phn & 53734.26454\phn & 2.6\phn & 0.142\phn \\
\phn & \phn & \phn & \phn & 3\phn & 53756.09692\phn & 3.1\phn & 0.134\phn \\
2\phn & \object[2MASS J05522624+3240446]{V827}\phn & 055226.24+324044.62\phn & 15.774\phn & 1\phn & 53731.14064\phn & 1.1\phn & 0.074\phn \\
\phn & \phn & \phn & \phn & 2\phn & 53745.26403\phn & 4.7\phn & 0.171\phn \\
3\phn & \object[Cl* NGC 2099    HGH V926]{V926}\phn & 055232.65+322208.16\phn & 16.348\phn & 1\phn & 53755.08741\phn & 2.2\phn & 0.046\phn \\
4\phn & \object[Cl* NGC 2099    HGH V985]{V985}\phn & 055236.13+323001.13\phn & 16.012\phn & 1\phn & 53734.15888\phn & 2.2\phn & 0.054\phn \\
\phn & \phn & \phn & \phn & 2\phn & 53756.24427\phn & 1.1\phn & 0.051\phn \\
5\phn & \object[Cl* NGC 2099    HGH V1018]{V1018}\phn & 055237.96+323415.64\phn & 16.569\phn & 1\phn & 53732.14053\phn & 1.5\phn & 0.019\phn \\
\phn & \phn & \phn & \phn & 2\phn & 53756.17928\phn & 1.8\phn & $<$ 0.030\phn \\
6\phn & \object[Cl* NGC 2099    HGH S29]{V1483}\phn & 055311.98+322353.58\phn & 17.717\phn & 1\phn & 53726.20141\phn & 0.4\phn & 0.208\phn \\
\phn & \phn & \phn & \phn & 2\phn & 53730.29695\phn & 0.7\phn & 0.146\phn \\
\phn & \phn & \phn & \phn & 3\phn & 53733.22359\phn & 1.6\phn & 0.184\phn \\
\phn & \phn & \phn & \phn & 4\phn & 53734.41413\phn & 0.4\phn & 0.159\phn \\
\phn & \phn & \phn & \phn & 5\phn & 53737.34213\phn & 1.2\phn & 0.161\phn \\
\phn & \phn & \phn & \phn & 6\phn & 53740.33536\phn & 0.9\phn & 0.124\phn \\
\phn & \phn & \phn & \phn & 7\phn & 53753.25941\phn & 0.7\phn & 0.203\phn \\
\phn & \phn & \phn & \phn & 8\phn & 53756.17828\phn & 1.2\phn & 0.157\phn \\
7\phn & V1731\phn & 055157.49+324534.90\phn & 15.823\phn & 1\phn & 53728.25931\phn & 3.3\phn & 0.233\phn \\
\phn & \phn & \phn & \phn & 2\phn & 53732.11914\phn & 2.5\phn & 0.282\phn \\
8\phn & V1748\phn & 055159.96+322433.64\phn & 14.846\phn & 1\phn & 53740.46821\phn & 0.5\phn & 0.012\phn \\
\phn & \phn & \phn & \phn & 2\phn & 53742.46911\phn & 0.5\phn & 0.015\phn \\
9\phn & V1757\phn & 055201.06+322408.79\phn & 17.189\phn & 1\phn & 53731.14064\phn & 0.4\phn & 0.022\phn \\
10\phn & V1889\phn & 055216.11+322822.95\phn & 14.971\phn & 1\phn & 53742.42521\phn & 1.6\phn & 0.100\phn \\
\phn & \phn & \phn & \phn & 2\phn & 53756.13038\phn & 3.8\phn & 0.082\phn \\
11\phn & V1892\phn & 055216.39+322609.00\phn & 19.328\phn & 1\phn & 53734.15888\phn & 2.0\phn & 0.178\phn \\
12\phn & V1909\phn & 055218.62+324100.16\phn & 13.281\phn & 1\phn & 53731.34461\phn & 0.0\phn & 0.019\phn \\
\phn & \phn & \phn & \phn & 2\phn & 53753.15826\phn & 2.0\phn & 0.291\phn \\
13\phn & V1915\phn & 055218.84+324434.08\phn & 21.189\phn & 1\phn & 53755.21194\phn & 1.9\phn & 0.273\phn \\
14\phn & V2213\phn & 055301.96+323744.91\phn & 16.433\phn & 1\phn & 53725.17392\phn & 2.0\phn & 0.230\phn \\
\phn & \phn & \phn & \phn & 2\phn & 53733.15900\phn & 0.5\phn & 0.025\phn \\
15\phn & \object[Cl* NGC 2099    HGH   80009]{V2214}\phn & 055302.02+322327.62\phn & 15.094\phn & 1\phn & 53732.11914\phn & 2.6\phn & 0.091\phn \\
\phn & \phn & \phn & \phn & 2\phn & 53735.16000\phn & 1.9\phn & 0.087\phn \\
\phn & \phn & \phn & \phn & 3\phn & 53754.14737\phn & 0.8\phn & 0.051\phn \\

\enddata
\tablenotetext{a}{$t_\mathrm{0}$ is the first time of dimming points that detected by CPA (in units of MJD).}
\tablenotetext{b}{$\Delta t$ is the duration between the start time of eclipse and the end (in units of hours).}
\tablenotetext{c}{$d$ is the depth of observed dimming points.}
\end{deluxetable}

% Section 5.1.3
\subsubsection{New EB candidates with eclipsing-like features}
With our CPA analysis, we discovered 15 new eclipsing binary candidates that have eclipsing-like or transit-like features undergoing eclipse, eclipse-ingress, or eclipse-egress phases.  Table \ref{Tab3} summarize observed eclipsing features of the light curves, which are missed by the standard analysis software.  Their orbital periods are not well constrained.  The light curve of each event can be described by the first time of dimming points $t_\mathrm{0}$, the time duration between the start time of eclipse and the end time of eclipse $\Delta t$, and the depth of observed dimming points $d$.  Figure \ref{Fig11} illustrates examples of light curves of this category.

\paragraph{V791} This detached EB system has deep primary and secondary eclipses (12--13\% and 1\%, respectively), and has out-of-eclipse variability at the 2\% level with period $\sim$4.0631 days.  After pre-whitening with the rotational modulation, our analysis indicates possible orbital period $P_\mathrm{orb}$ = 2.4277 days.

\paragraph{V926} This object was known to be a long-period (16.5777 days) variable star.  Our LS analysis supports the conclusion of previous study, but it has sinusoidal-shaped periodic signal in our light curve.  The LS periodogram shows two power peaks with periods similar to previous one (rank 1 = 15.7291 days) and another shorter than one day (rank 2 = 0.9376 days).  One sudden eclipse event was caught during its return to quiescence brightness.

% Figure 12
\begin{figure}[!t]
\begin{center}
  \includegraphics[width=1.0\linewidth, angle=0]{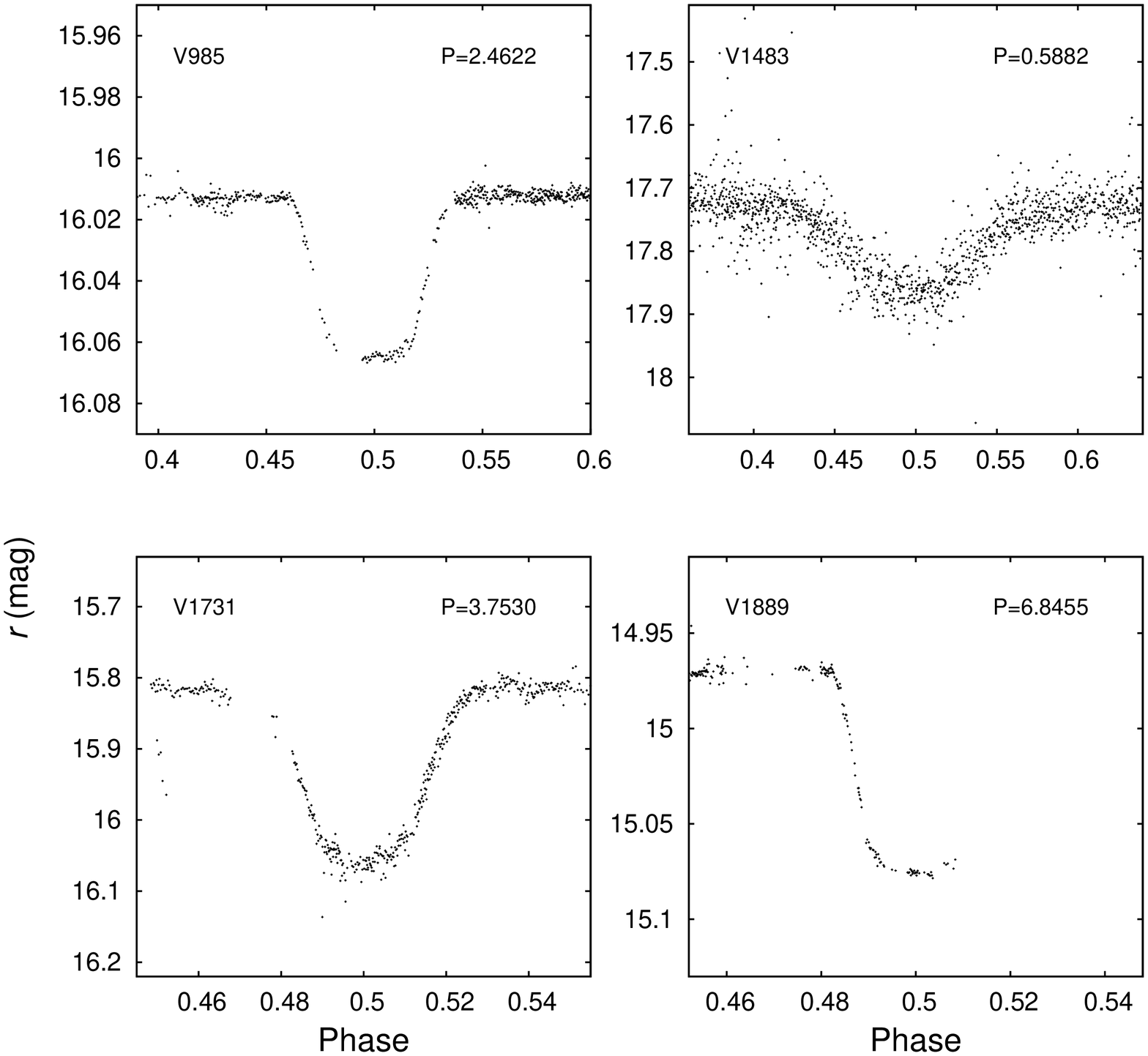}  
  \caption{Phased light curves of V985, V1483, V1731, and V1889.  The mid-eclipse is set to the orbital phase of 0.5.  The estimated orbital period is given in the upper right of each panel.} 
  \label{Fig12}
\end{center}
\end{figure}

\paragraph{V1018} We caught two eclipses, but the star seems to vary all the time in a strange manner. %RS Canum Venaticorum variable?

% Section 5.1.4
\subsubsection{Grazing EB candidates}
By using the Plavchan algorithm \citep{pla08} provided in NASA Exoplanet Archive Periodogram Service\footnote{NASA Exoplanet Archive Periodogram Service is available at \url{http://exoplanetarchive.ipac.caltech.edu/cgi-bin/Periodogram/nph-simpleupload}.}, we identified two types of systems which can be confused with a transiting planetary system.  One type is connected with the grazing EBs, and it shows V- or U-shape eclipses.  The other type is the eclipsing binary systems with a large primary star, in which case the eclipse produces a flat bottomed light curve with small deeming.  Figure \ref{Fig12} shows the phased light curves of actual cases of these contaminants.  The observed shapes of the eclipsing events resemble transit-like signals, but depths of mid-eclipse are more than a few \% of the total light (V985: 5\%; V1483: 14\%; V1731: 24\%; V1889: 10\%).

% Section 5.2
\subsection{Multiperiodic variable stars}
Multi-periodicity is a common feature found in nearly all type of pulsating stars such as $\delta$ Scuti stars, RR Lyraes, and Cepheids.  Our catalog contains many examples of multiperiodic variable stars which deserve in-depth analysis.

\begin{figure}[!t]
	\includegraphics[width=\linewidth, angle=0]{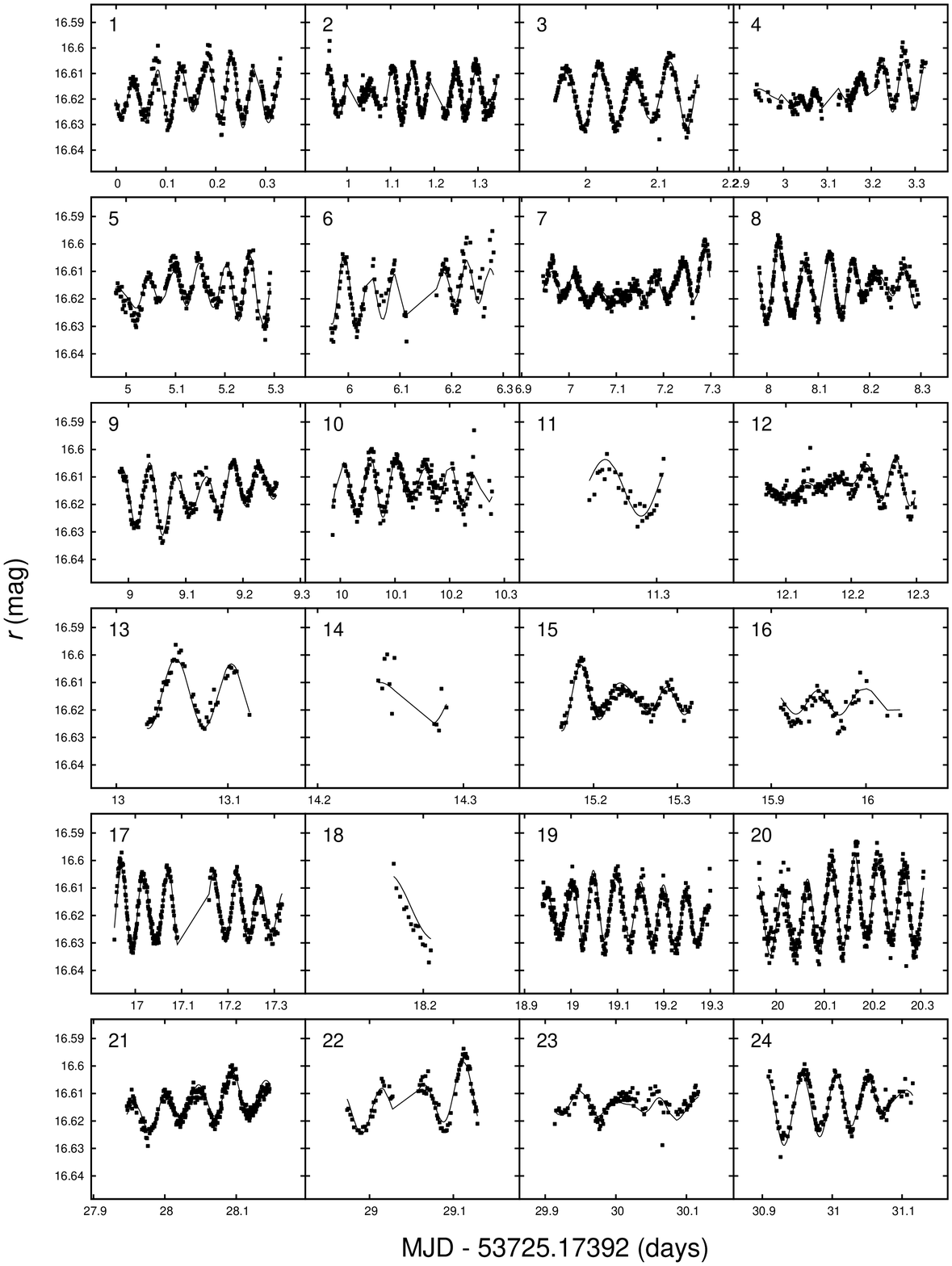}
    \caption{Example light curve of multiperiodic variable V1898.  The fit of the 16-frequency solution presented in the Table \ref{TabA2} is shown as a solid curve.  The full amplitude of this variation is about 0.04 mag in $r$-band light curve.}
 \label{Fig13}
\end{figure}

% Section 5.2.1
We performed multiple-frequency analysis of 92 short-periodic variables with the program {\sc PERIOD04} \citep{len05}, looking for the presence of multi-periodicity.  With an iterative pre-whitening procedure, discrete Fourier transformations are calculated until there is no meaningful change in the fit residuals.  Each light curve was fitted using an equation of the form:
\begin{equation}
y(t) = A_{0} + \sum_{j=1}^{n} A_{j} \cos(2\pi f_{j} t - \phi_{j})
\end{equation} where $A_{0}$ is a mean level and $f_{j}$, $A_{j}$, and $\phi_{j}$ are the frequency, amplitude, and phase for each successive peak found in the amplitude spectrum.  We adopt a conservative approach in selecting the statistical significant peaks from the amplitude spectrum.  In general, a signal/noise (S/N) amplitude ratio of 4.0 is a good criterion for independent frequencies, equivalent to 99.9\% certainty of variability \citep{bre93, kus97}.  For combination modes with known values, the S/N amplitude ratio of $\sim$3.5 is used which is equivalent to 90\% certainty.  After removal of all significant peaks in the amplitude spectrum, the noise level was estimated on the residual amplitude spectrum over 5 c/d boxes around each frequency.  The resulting light curve of one example multiperiodic variable is shown in Figure \ref{Fig13}.  The multi-period model is in excellent agreement with the observed light curve.

% Figure 13

Table \ref{TabA2} summarizes all candidate frequencies identified in 92 multiperiodic light curves with their fitted Fourier parameters.  All the detected frequencies are classified as independent modes ($f_{1}$, $f_{2}$, $f_{3}$, $\cdots$), combination modes ($af_{i} \pm bf_{j}$), and harmonic mode ($hf_{j}$).  

% Section 5.3
\subsection{Aperiodic variability}
\begin{figure}[!t]
  \includegraphics[width=1.0\linewidth, angle=0]{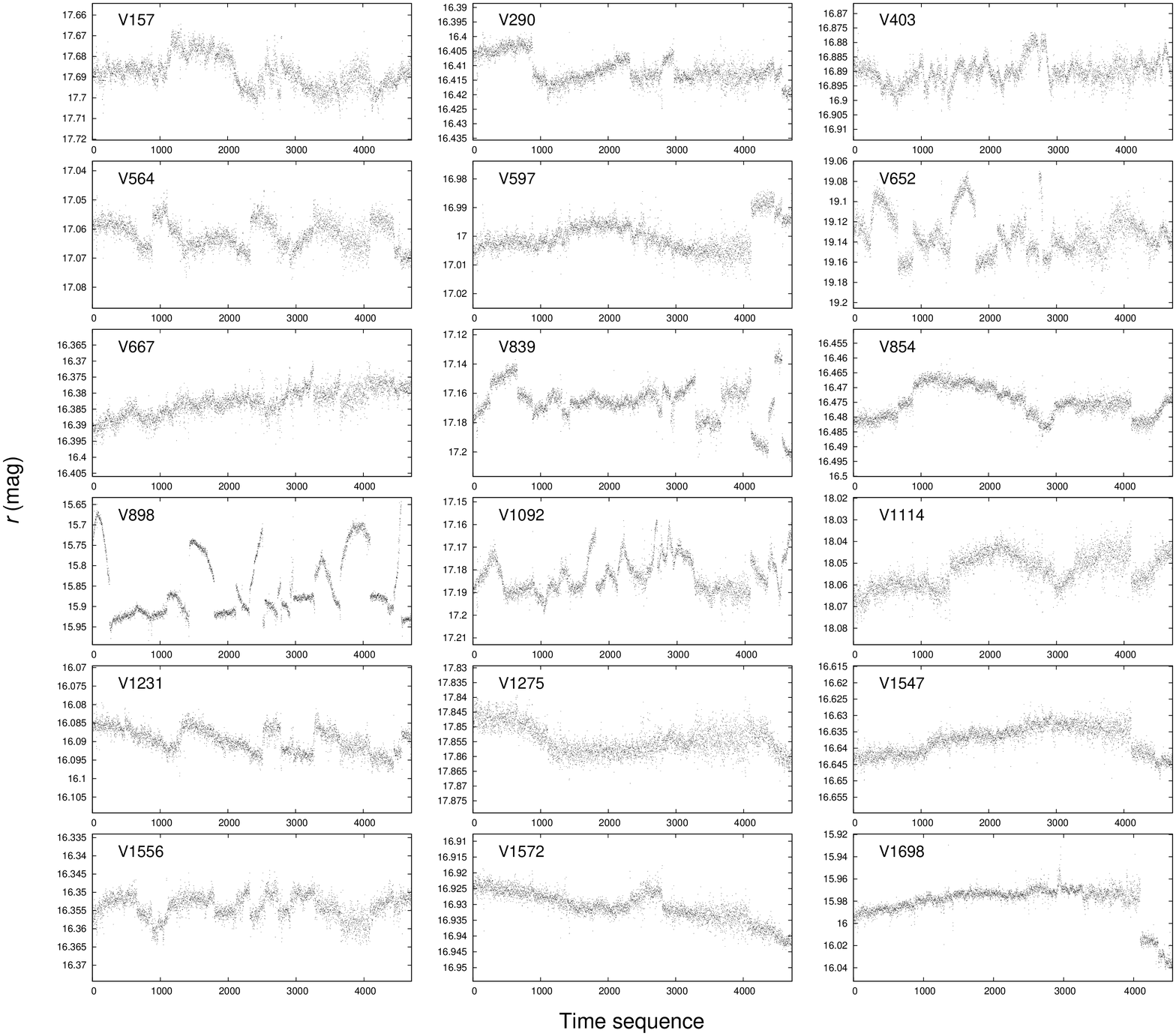}  
  \caption{Examples of aperiodic light curves showing variations about the mean, in addition to a relatively long-term variability on timescales similar to the total observing time span ($\sim$31 days).}
  \label{Fig14}
\end{figure}

By using the method described in Section 3.1, we detected long-term aperiodic variability of 132 variable candidates with time-scales similar to the total observing time span ($\sim$31 days).  We visually inspected each light curve to confirm its variable nature, and this result is illustrated in terms of time sequence (Figure \ref{Fig14}).  Due to the lack of periodicity, the light curves of aperiodic variables are simply characterized by the level of photometric variability ($\Delta m$), the dispersion about the mean level ($\sigma_{m}$), and the reduced chi-square assuming a constant median value for the light curve ($\chi_{\nu}^{2}$).
\begin{equation}
\chi_{\nu}^{2} = \frac{1}{N - 1}\sum_{i}^{N}\frac{(x_{i} - \bar{x})^{2}}{\sigma_{i}^{2}}
\end{equation} where $\sigma_{i}$ is the photometric uncertainty of each data point $x_{i}$, and $N$ is the total number of measurements.  The full parameters are described in Table \ref{TabA3}.  These variables show relatively low levels of variability with $\Delta m$ $<$ 0.03 mag (60\%) and $\Delta m$ $<$ 0.05 mag (80\%), respectively.

% Figure 14

% Figure 15
\begin{figure}[!t]
  \includegraphics[width=\linewidth, angle=0]{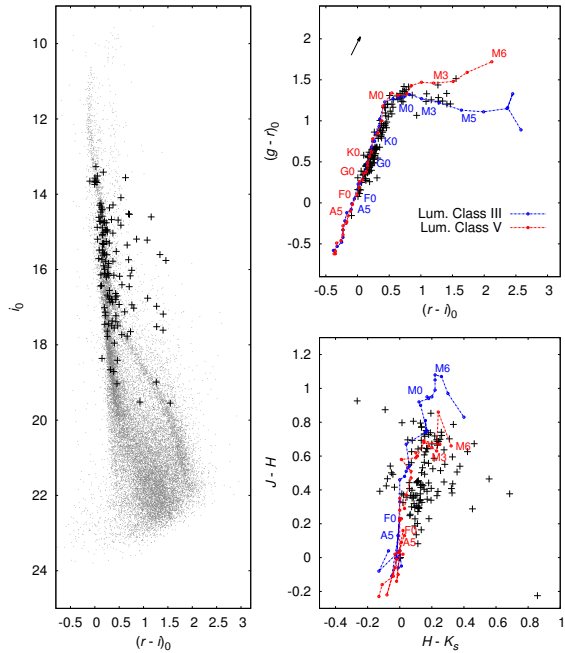}  
  \caption{Color-magnitude diagram and color-color diagrams of aperiodic variable candidates (black points).  The color index distributions follow the locus of normal stars taken from \citet{cov07}.  The two lines show the location of the median stellar locus of dwarf (V) and giant (III) stars having solar-metallicity.  The cluster\rq{s} extinction is taken into account for SDSS color-color diagram.  In order to derive $A_{\lambda}/A_{V}$ values for SDSS photometry system, we used the extinction coefficients compiled in \citet{gir04}.}
  \label{Fig15}
\end{figure}

We used the combined SDSS and 2MASS photometry to identify basic properties of these sources.  Since color ranges are much smaller in the near-infrared color-color diagram, the stellar locus is not as well defined as in optical color-color diagram (Figure \ref{Fig15}).  We infer spectral types from \citet{cov07}, in which stars in this sample cover spectral types from approximately A5 to M5 and can be either main-sequence and giant stars.  The CMD indicates that many of these aperiodic stars are field stars not associated with M37.

The origin of aperiodic nature is difficult to explain.  Possible speculations include that (i) some of these may be cases where long-period field giants are detected as low-amplitude photometric variables on timescales of weeks or longer (e.g., \citealt{hen00,cia11}), (ii) some of them may be semi-regular variables exhibiting both fast and slow variabilities (e.g., \citealt{and13}),  and (iii) as mentioned in \citet{har08b}, the rapid amplitude variation may be flickering, nova-like outbursts seen in many symbiotic variables (e.g., \citealt{gro09,gro13}).  Further observations of much longer duration will be necessary to properly characterize these aperiodic variables.

% Section 5.4
\subsection{Flare stars}
Earlier analysis of M37 data (Hartman et al. 2008a) could not address flare events properly because many data points were removed as outliers.  Our new pipeline, with nearly 100\% data utilization, offers opportunity to investigate flare stars more thoroughly.

\begin{figure}[!t]
  \includegraphics[width=1.0\linewidth, angle=0]{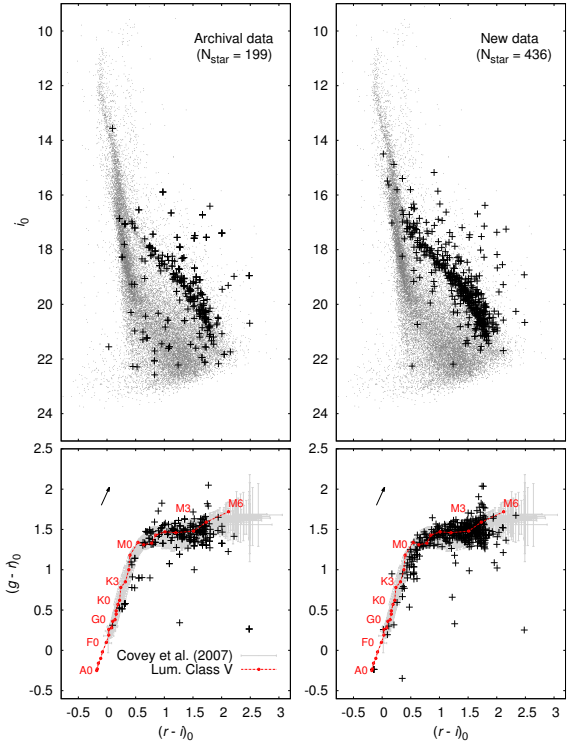}  
  \caption{Color-magnitude and color-color diagrams of flaring stars, detected in archival (left panels) and our new data (right panels), respectively.  The red lines represent the main-sequence stellar locus with solar metallicity \citep{cov07}.  Most of these stars can be regarded as late-type main sequence stars (K4--M6) in the M37 open cluster.}
  \label{Fig16}
\end{figure}

By CPA analysis, we found that 436 stars (625 flare events) show flaring activity characterized by a rapid flux increase and a slower decay.  As expected, the location of these flaring stars on color-magnitude and color-color diagrams suggest that most of them are late-type main-sequence stars (K4--M6) in M37 open cluster (Figure \ref{Fig16}).  Flares are also observed from earlier spectral types (A0--K3), but these are only a few and probably belong to field stars.

% Figure 16

The underlying mechanism of flares in stars cooler than about F5 is essentially the same as in the Sun.  It is believed that these cool stars can generate a magnetic field through a dynamos at the interface between the radiative core and the convective envelope \citep{benz10}.  Thus, flaring is an inevitable result of strong magnetic field.  But flares in hot stars, especially in A-type stars, are not easy to explain in the context of dynamo process because these stars should not generate strong magnetic fields in their thin convection layer and are incapable of storing larger energy in that region \citep{bal12}.  

The observed flare signature of one A-type star (V72) might be the result of binarity.  If an A-type main sequence star has an active M dwarf secondary star, such binary star would look like a flare-generating A-type star.

\begin{figure}[!t]
  \includegraphics[width=\linewidth, angle=0]{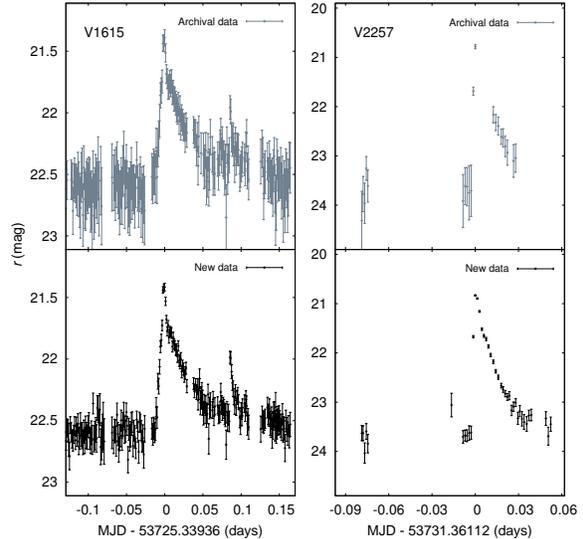}         
  \caption{Comparison between our and archival light curves of the two flare events.  As shown in Paper I, new analysis provides more improved $r$-band data in terms of both photometric accuracy and data recovery rate.}
  \label{Fig17}
\end{figure}

\begin{figure*}[!t]
\centering
   \subfloat[]{\includegraphics[width=0.83\linewidth, angle=0]{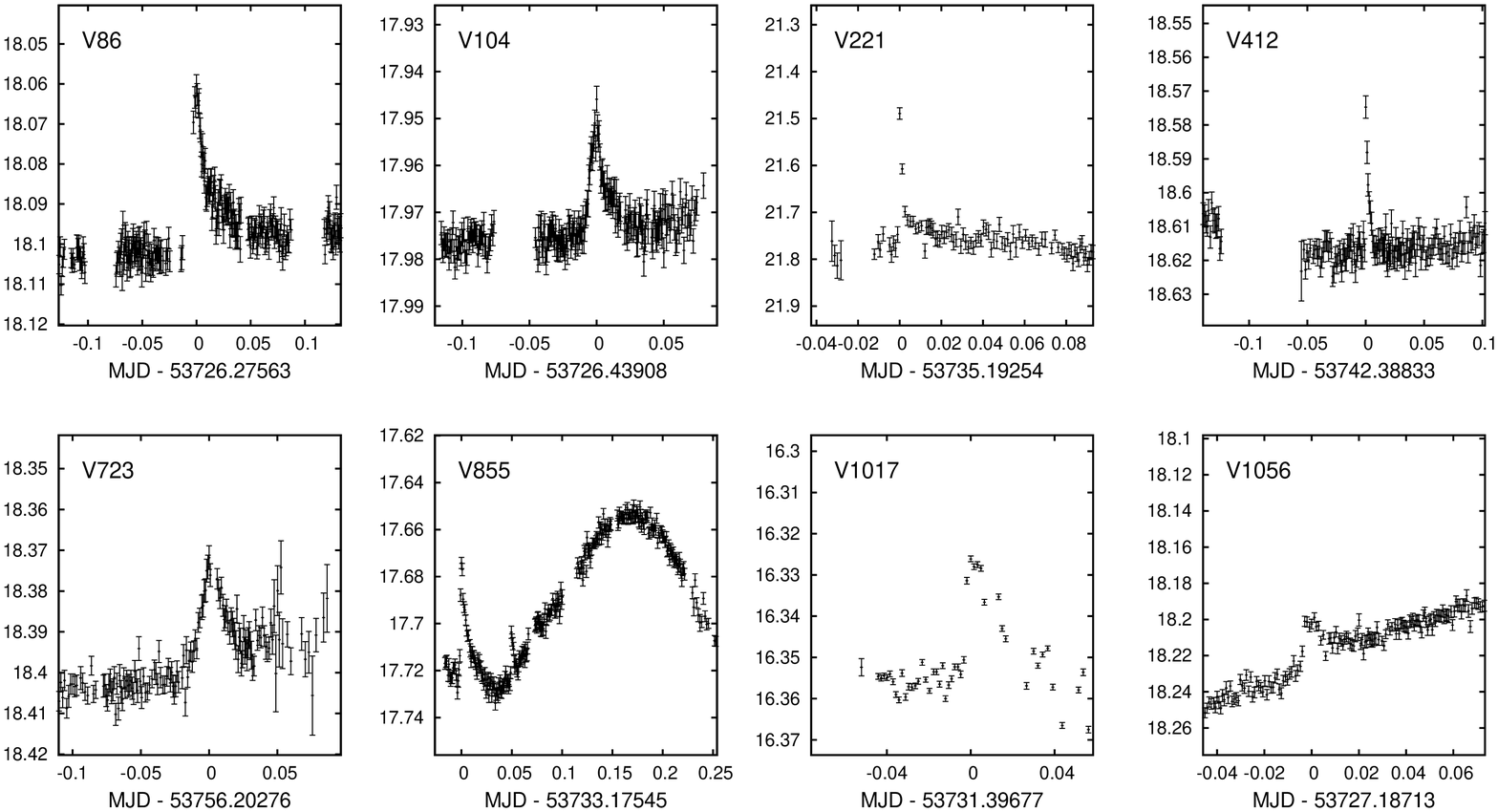}}\vspace*{-1.3em}
   \subfloat[]{\includegraphics[width=0.83\linewidth, angle=0]{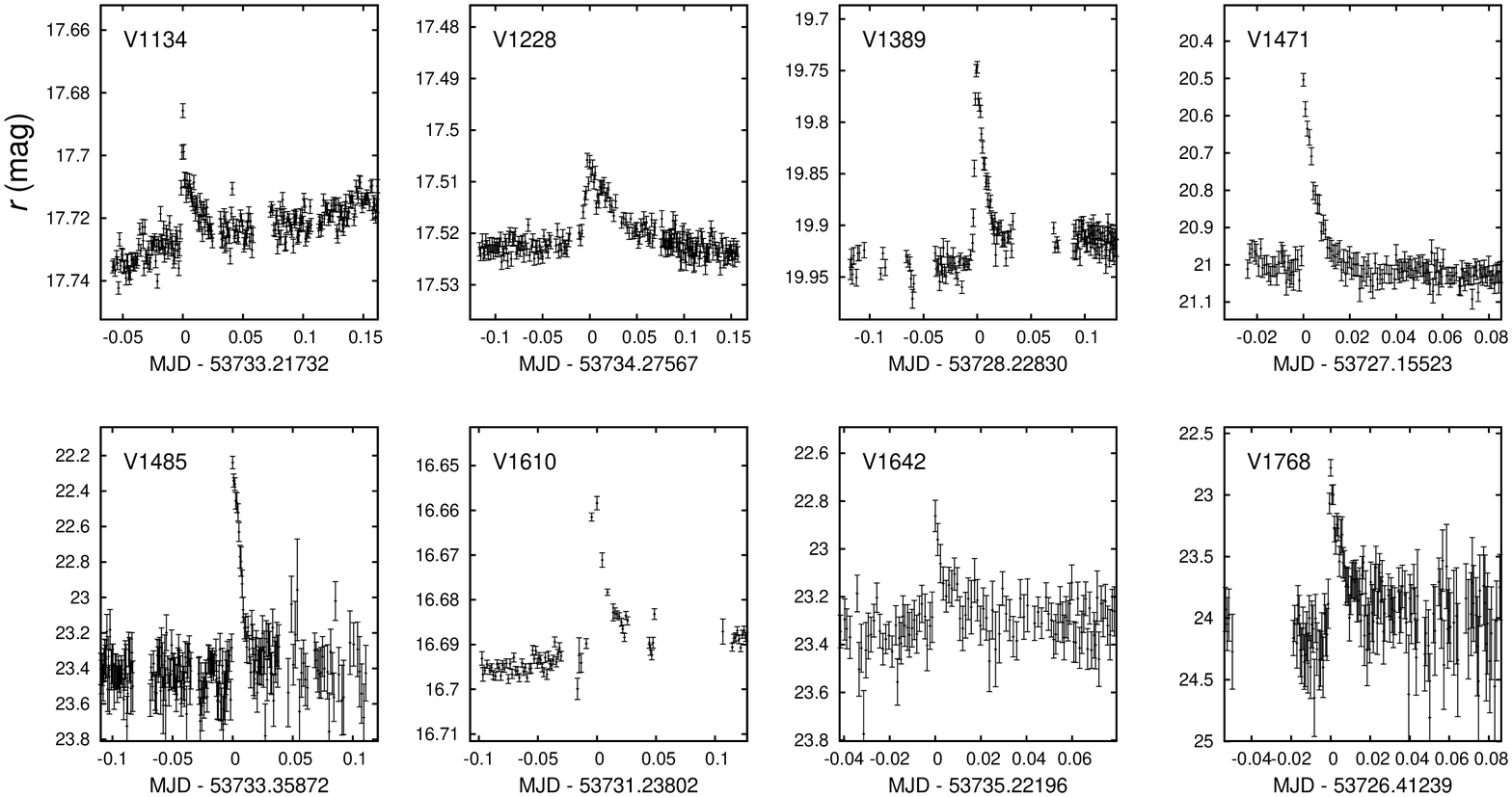}}\vspace*{-1.3em}
   \subfloat[]{\includegraphics[width=0.83\linewidth, angle=0]{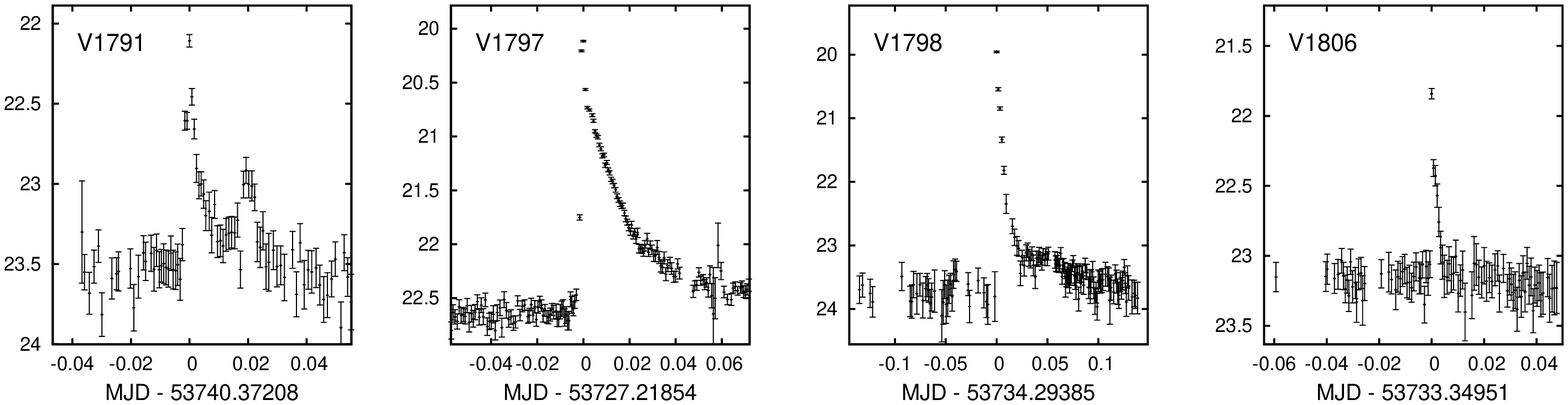}}\vspace*{-0.8em}
   \caption{Various light curves of discovered flare events selected by color cut ($(g-r)_{0} < 1$).  Most of them show a sharp rise and subsequent exponential decay.}
  \label{Fig18}
\end{figure*}

The number of flare events is increased by a factor of two in the new analysis because our new calibration procedure provides more accurate $r$-band light curves (Figure \ref{Fig17}).  In Table \ref{TabA4} and Figure \ref{Fig18}, we list the parameters of flare events discovered by our CPA analysis.  They have a broad range of peak amplitude ($\Delta m_\mathrm{peak}$ = $0.006\sim3.719$ mag) and typical durations of a few minutes to several hours, irrespective of their spectral type.  More detailed analysis of flare statistics and its relation to magnetic activity and stellar rotation will be presented in subsequent papers.

% Figure 17

% Figure 18

% Section 6
\section{Conclusion}
We cataloged and investigated  a variety of photometric variables in the field of M37 using our new light curves, which were obtained from archival imaging data of one-month long exo-planet transit survey with the 6.5-m MMT telescope.  This new dataset allows a rare opportunity to explore different types of variability of short ($\sim$minutes) and long ($\sim$one-month) time-scales simultaneously.  

We find 61 eclipsing binary systems (EB), 92 multiperiodic variable stars (mp), 132 aperiodic variables (A), and 436 flare stars (F), as well as several hundreds of rotating variables (R).  Our work has increased the number of known variable stars in this field by 60\%.  The new variable catalog has been made available on the web (\url{http://stardb.yonsei.ac.kr/}).

% Acknowledgments
\acknowledgements
We would like to thank the anonymous referee for helpful suggestion.  This research was supported by Basic Science Research Program of the National Research Foundation of Korea (2011-0030875).  Y.-I.B. acknowledges the support from KASI-Yonsei DRC program of Korea Research Council of Fundamental Science and Technology (DRC-12-2-KASI).

%"This work was supported by the Brain Korea 21 Plus Program in 2015 (21A20131500002)" 

% References

\clearpage

% Appendix
\renewcommand{\thetable}{A\arabic{table}}
\setcounter{table}{0}  

% Table A1
\begin{deluxetable}{ccrrrrrrrrrrrrc}
\tabletypesize{\scriptsize}
\rotate
\tablecaption{New catalog of 2306 variable stars in M37 field\label{TabA1}}
\tablewidth{0pt}
\tablehead{
\colhead{} & \colhead{StarID} & \colhead{Period\tablenotemark{a}} & \colhead{Amp\tablenotemark{a}} &
\colhead{$u$\tablenotemark{b}} & \colhead{$r$\tablenotemark{c}} & \colhead{$g-r$\tablenotemark{c}} & \colhead{$r-i$\tablenotemark{c}} & \colhead{$z$\tablenotemark{b}} & \colhead{$V$\tablenotemark{d}} & \colhead{$B-V$\tablenotemark{d}} & \colhead{$J$\tablenotemark{e}} & \colhead{$J-H$\tablenotemark{e}} & \colhead{$H-K_{s}$\tablenotemark{e}} & \colhead{} \\
\colhead{VarID} & \colhead{(J2000)} & \colhead{(days)} & \colhead{(mag)} &
\colhead{(mag)} & \colhead{(mag)} & \colhead{(mag)} & \colhead{(mag)} & \colhead{(mag)} & \colhead{(mag)} & \colhead{(mag)} & \colhead{(mag)} & \colhead{(mag)} & \colhead{(mag)} & \colhead{VarType\tablenotemark{f}}
}
\startdata
\object[V* V541 Aur]{V1}\phn & 055220.39+323319.65\phn & \nodata\phn & \nodata\phn & \nodata\phn & 13.677\phn & 0.285\phn & 0.191\phn & \nodata\phn & 13.406\phn & 0.609\phn & 12.447\phn & 0.270\phn & 0.063\phn & EB\phn\\
\object[V* V540 Aur]{V2}\phn & 055216.58+322815.32\phn & \nodata\phn & \nodata\phn & 16.792\phn & 14.840\phn & 0.564\phn & 0.290\phn & 14.471\phn & 14.653\phn & 0.688\phn & 13.539\phn & 0.287\phn & 0.134\phn & EB\phn\\
\object[V* V542 Aur]{V3}\phn & 055233.01+323241.91\phn & 0.4225\phn & 0.320\phn & 17.982\phn & 15.789\phn & 0.560\phn & 0.329\phn & 15.284\phn & 16.051\phn & 0.728\phn & 14.267\phn & 0.355\phn & 0.012\phn & EB\phn\\
\object[V* V544 Aur]{V4}\phn & 055253.26+323301.47\phn & 0.5582\phn & 0.235\phn & 18.079\phn & 15.634\phn & 0.762\phn & 0.477\phn & 15.093\phn & 15.948\phn & 0.938\phn & 13.951\phn & 0.497\phn & 0.101\phn & EB\phn\\
\object[V* V545 Aur]{V5}\phn & 055300.63+322450.81\phn & 0.2788\phn & 0.310\phn & 18.036\phn & 15.915\phn & 0.459\phn & 0.481\phn & 15.235\phn & 16.184\phn & 0.585\phn & 14.421\phn & 0.359\phn & 0.084\phn & mp\phn\\
\object[V* V539 Aur]{V6}\phn & 055150.53+323234.70\phn & 0.1098\phn & 0.343\phn & 18.218\phn & 15.973\phn & 0.591\phn & 0.173\phn & 15.622\phn & \nodata\phn & \nodata\phn & 14.739\phn & 0.342\phn & -0.009\phn & mp\phn\\
\object[Cl* NGC 2099    HGH V7]{V7}\phn & 055239.09+323631.27\phn & 0.3577\phn & 0.433\phn & 19.858\phn & 17.252\phn & 1.039\phn & 0.056\phn & 17.065\phn & 17.890\phn & 0.815\phn & 15.779\phn & 0.446\phn & 0.196\phn & EB\phn\\
\object[Cl* NGC 2099     KV       1]{V8}\phn & 055234.32+323218.70\phn & 0.1195\phn & \nodata\phn & \nodata\phn & \nodata\phn & \nodata\phn & \nodata\phn & \nodata\phn & \nodata\phn & \nodata\phn & \nodata\phn & \nodata\phn & \nodata\phn & U\phn\\
\object[2MASS J05521491+3224415]{V9}\phn & 055214.91+322441.35\phn & 0.0909\phn & \nodata\phn & 14.873\phn & 13.233\phn & 0.137\phn & 0.128\phn & 13.114\phn & 13.298\phn & 0.370\phn & 12.348\phn & 0.171\phn & 0.042\phn & P\phn\\
\object[2MASS J05520049+3236481]{V10}\phn & 055200.49+323648.19\phn & 0.9432\phn & 0.075\phn & 16.844\phn & 14.151\phn & 0.928\phn & 0.436\phn & 13.493\phn & 14.762\phn & 0.531\phn & 12.356\phn & 0.567\phn & 0.131\phn & P\phn\\
\enddata
\tablecomments{Table \ref{TabA1} is published in its entirety in the electronic edition of {\it Astronomical Journal}.  A portion is shown here for guidance regarding its form and content.}
\tablenotetext{a} {Dominant period and its full amplitude.  All blank fields are non-periodic variables.  In cases where the periodicity analysis is affected by a large scatter of light curves due to photometric bias or source blending, we used the already known period in that source from the previous catalog \citep{har08b}.}
\tablenotetext{b} {$u$ and $z$ magnitudes from the Sloan Digital Sky Survey (SDSS) DR7 \citep{aba09}.}
\tablenotetext{c} {$g$, $r$, and $i$ magnitudes from \citet{har08a} and SDSS DR7 catalog.}
\tablenotetext{d} {$B$ and $V$ photometry from \citet{kal01}.}
\tablenotetext{e} {$J$, $H$, and $K_s$ photometry from the Two Micron All Sky Survey (2MASS) Point Source Catalog \citep{cut03}.}
\tablenotetext{f} {Type of variability. The following types of variability were assigned: F = flare stars, P = pulsating variables, mp = multiperiodic variables, R = rotating variables, EB = eclipsing binary systems, A = aperiodic variables, and : (or var:) = variable candidates.  We also use the two cases: nonvar = non-variables listed in Table \ref{Tab1}, U = These variable objects fell on chip gaps and were not observed.}
\end{deluxetable}
\clearpage

% Table A2
\begin{deluxetable}{rccrrcrrrl}
\tabletypesize{\footnotesize}
\tablecaption{Candidate frequencies identified in 92 multiperiodic light curves \label{TabA2}}
\tablewidth{0pt}
\tablehead{
\colhead{} & \colhead{} & \colhead{Star} & \colhead{$r$} & \colhead{Frequency} & \colhead{Frequency} & \colhead{$A_{r}$\tablenotemark{a}} & \colhead{$\phi$} & \colhead{} & \colhead{} \\
\colhead{No.} & \colhead{VarID} & \colhead{(J2000)} & \colhead{(mag)} & \colhead{(cd$^{-1}$)} & \colhead{name} & \colhead{(mmag)} & \colhead{(deg)} & \colhead{S/N\tablenotemark{b}} & \colhead{Note\tablenotemark{c}}
}
\startdata
1\phn & \object[V* V545 Aur]{V5}\phn & 055616.79+322514.20\phn & 15.915\phn & 3.5872\phn & f$_{1}$\phn & 310.1\phn & 0.3331\phn & 68.9\phn & \phn\\ 
\phn & \phn & \phn & \phn & 7.1768\phn & f$_{2}$(=2f$_{1}$)\phn & 35.4\phn & 0.2906\phn & 11.3\phn & \phn\\ 
\phn & \phn & \phn & \phn & 10.7619\phn & f$_{3}$(=3f$_{1}$)\phn & 22.8\phn & 0.0418\phn & 10.3\phn & \phn\\ 
\phn & \phn & \phn & \phn & 5.6768\phn & f$_{4}$\phn & 19.5\phn & 0.2369\phn & 5.7\phn & \phn\\ 
\phn & \phn & \phn & \phn & 14.3473\phn & f$_{5}$\phn & 13.8\phn & 0.9990\phn & 7.1\phn & \phn\\ 
2\phn & \object[V* V539 Aur]{V6}\phn & 055506.89+323303.20\phn & 15.973\phn & 9.1040\phn & f$_{1}$(=f$_{3}$-f$_{2}$)\phn & 343.4\phn & 0.5433\phn & 172.7\phn & \phn\\ 
\phn & \phn & \phn & \phn & 18.2081\phn & f$_{2}$(=f$_{3}$-f$_{1}$)\phn & 89.9\phn & 0.2201\phn & 53.6\phn & \phn\\ 
\phn & \phn & \phn & \phn & 27.3122\phn & f$_{3}$(=f$_{1}$+f$_{2}$)\phn & 27.6\phn & 0.9237\phn & 19.7\phn & \phn\\ 
\phn & \phn & \phn & \phn & 8.7545\phn & f$_{4}$\phn & 15.4\phn & 0.5801\phn & 8.1\phn & \phn\\ 
\phn & \phn & \phn & \phn & 11.9319\phn & f$_{5}$\phn & 14.5\phn & 0.5557\phn & 6.3\phn & \phn\\ 
\phn & \phn & \phn & \phn & 36.4178\phn & f$_{6}$(=2f$_{2}$)\phn & 10.1\phn & 0.0527\phn & 10.4\phn & \phn\\ 
\phn & \phn & \phn & \phn & 14.9868\phn & f$_{7}$\phn & 8.8\phn & 0.9887\phn & 4.5\phn & \phn\\ 
\phn & \phn & \phn & \phn & 16.8695\phn & f$_{8}$\phn & 7.9\phn & 0.7280\phn & 4.5\phn & \phn\\ 
3\phn & \object[2MASS J05520338+3235136]{V11}\phn & 055519.81+323541.23\phn & 15.245\phn & 8.4261\phn & f$_{1}$\phn & 79.0\phn & 0.9052\phn & 84.8\phn & \phn\\ 
\phn & \phn & \phn & \phn & 16.8561\phn & f$_{2}$(=2f$_{1}$)\phn & 7.6\phn & 0.6488\phn & 8.3\phn & \phn\\ 
\phn & \phn & \phn & \phn & 17.7044\phn & f$_{3}$\phn & 8.0\phn & 0.5895\phn & 9.1\phn & \phn\\ 
\phn & \phn & \phn & \phn & 16.4751\phn & f$_{4}$\phn & 6.9\phn & 0.5617\phn & 7.3\phn & \phn\\ 
\phn & \phn & \phn & \phn & 8.8787\phn & f$_{5}$\phn & 4.9\phn & 0.4440\phn & 5.2\phn & \phn\\ 
\phn & \phn & \phn & \phn & 12.0801\phn & f$_{6}$\phn & 4.8\phn & 0.8357\phn & 5.0\phn & \phn\\ 
\phn & \phn & \phn & \phn & 9.2640\phn & f$_{7}$\phn & 5.9\phn & 0.6453\phn & 6.4\phn & \phn\\ 
\phn & \phn & \phn & \phn & 10.6455\phn & f$_{8}$\phn & 5.4\phn & 0.0597\phn & 6.0\phn & \phn\\ 
\phn & \phn & \phn & \phn & 13.5604\phn & f$_{9}$\phn & 4.8\phn & 0.8005\phn & 4.9\phn & \phn\\ 
\phn & \phn & \phn & \phn & 16.2578\phn & f$_{10}$\phn & 5.1\phn & 0.6379\phn & 5.3\phn & \phn\\ 
\phn & \phn & \phn & \phn & 8.0470\phn & f$_{11}$\phn & 4.5\phn & 0.8268\phn & 4.7\phn & \phn\\ 
\phn & \phn & \phn & \phn & 2.3825\phn & f$_{12}$\phn & 5.2\phn & 0.7274\phn & 3.6\phn & \phn\\ 
\phn & \phn & \phn & \phn & 4.3812\phn & f$_{13}$\phn & 4.3\phn & 0.3942\phn & 4.2\phn & \phn\\ 
\phn & \phn & \phn & \phn & 14.9483\phn & f$_{14}$\phn & 3.9\phn & 0.8404\phn & 4.0\phn & \phn\\ 
4\phn & \object[2MASS J05524469+3230167]{V12}\phn & 055600.99+323041.25\phn & 15.873\phn & 12.7176\phn & f$_{1}$\phn & 14.4\phn & 0.0994\phn & 8.1\phn & \phn\\ 
\phn & \phn & \phn & \phn & 7.8896\phn & f$_{2}$\phn & 13.8\phn & 0.4901\phn & 6.0\phn & \phn\\ 
\phn & \phn & \phn & \phn & 12.7698\phn & f$_{3}$\phn & 10.5\phn & 0.0175\phn & 6.0\phn & close to f$_{1}$\phn\\ 
\phn & \phn & \phn & \phn & 14.4564\phn & f$_{4}$\phn & 9.6\phn & 0.0884\phn & 6.5\phn & \phn\\ 
5\phn & \object[2MASS J05521116+3225159]{V13}\phn & 055527.32+322543.01\phn & 16.511\phn & 10.0341\phn & f$_{1}$(=f$_{12}$-f$_{2}$)\phn & 23.0\phn & 0.4259\phn & 137.9\phn & \phn\\ 
\phn & \phn & \phn & \phn & 10.4706\phn & f$_{2}$(=f$_{12}$-f$_{1}$)\phn & 13.0\phn & 0.8171\phn & 77.9\phn & \phn\\ 
\phn & \phn & \phn & \phn & 17.1389\phn & f$_{3}$\phn & 8.7\phn & 0.9050\phn & 51.9\phn & \phn\\ 
\phn & \phn & \phn & \phn & 14.0376\phn & f$_{4}$\phn & 6.1\phn & 0.6801\phn & 33.6\phn & \phn\\ 
\phn & \phn & \phn & \phn & 13.0968\phn & f$_{5}$\phn & 3.7\phn & 0.1837\phn & 20.9\phn & \phn\\ 
\phn & \phn & \phn & \phn & 12.8019\phn & f$_{6}$\phn & 3.1\phn & 0.2056\phn & 17.4\phn & \phn\\ 
\phn & \phn & \phn & \phn & 15.0841\phn & f$_{7}$\phn & 2.1\phn & 0.7887\phn & 11.6\phn & \phn\\ 
\phn & \phn & \phn & \phn & 13.6263\phn & f$_{8}$\phn & 1.3\phn & 0.1260\phn & 7.1\phn & \phn\\ 
\phn & \phn & \phn & \phn & 11.1091\phn & f$_{9}$\phn & 1.4\phn & 0.2454\phn & 8.7\phn & \phn\\ 
\phn & \phn & \phn & \phn & 27.1718\phn & f$_{10}$\phn & 1.0\phn & 0.7296\phn & 6.7\phn & \phn\\ 
\enddata
\tablecomments{Table \ref{TabA2} is published in its entirety in the electronic edition of {\it Astronomical Journal}.  A portion is shown here for guidance regarding its form and content.}
\tablenotetext{a}{We doubled amplitudes derived by PERIOD04 since this program gives half of the full amplitudes.}
\tablenotetext{b}{The signal-to-noise ratio (S/N) was calculated by {\sc PERIOD04}.  Each noise level was computed from the prewhitened periodogram as a running mean over boxes of 5 cd$^{-1}$ in frequency.}
\tablenotetext{c}{We note the frequency values at which the close pairs occur with a separation less than 0.1 cd$^{-1}$}
\end{deluxetable}
\clearpage

% Table A3
\begin{deluxetable}{rccrrrrrrc}
\tabletypesize{\footnotesize}
\tablecaption{Long-term 132 aperiodic variable candidates with timescales of $>$1 month \label{TabA3}}
\tablewidth{0pt}
\tablehead{
\colhead{} & \colhead{} & \colhead{Star} & \colhead{$r$} & \colhead{$\sigma_{m}$\tablenotemark{a}} & \colhead{$\Delta$$m$\tablenotemark{b}} & \colhead {} & \colhead{} & \colhead{}\\
\colhead{No.} & \colhead{VarID} & \colhead{(J2000)} & \colhead{(mag)} & \colhead{(mag)} & \colhead{(mag)} & \colhead {$\chi^{2}_{\nu}$\tablenotemark{c}} & \colhead{N$_\mathrm{LC}$} & \colhead{Note\tablenotemark{d}}
}
\startdata
1\phn & \object[Cl* NGC 2099    HGH V58]{V58}\phn & 055122.14+322942.40\phn & 16.742\phn & 0.009\phn & 0.096\phn & 443.5\phn & 4730\phn & \nodata\phn \\
2\phn & \object[Cl* NGC 2099    HGH V70]{V70}\phn & 055123.93+324357.22\phn & 15.837\phn & 0.003\phn & 0.017\phn & 14.0\phn & 4706\phn & \nodata\phn \\
3\phn & \object[Cl* NGC 2099    HGH V75]{V75}\phn & 055124.73+324312.36\phn & 18.360\phn & 0.004\phn & 0.028\phn & 1.8\phn & 4730\phn & 13.260\phn \\
4\phn & \object[Cl* NGC 2099    HGH V87]{V87}\phn & 055125.42+323600.45\phn & 17.564\phn & 0.002\phn & 0.017\phn & 1.6\phn & 4730\phn & 21.199\phn \\
5\phn & \object[Cl* NGC 2099    HGH V107]{V107}\phn & 055127.04+323225.57\phn & 15.458\phn & 0.002\phn & 0.010\phn & 7.7\phn & 4592\phn & \nodata\phn \\
6\phn & \object[Cl* NGC 2099    HGH V114]{V114}\phn & 055128.19+322931.14\phn & 17.854\phn & 0.004\phn & 0.029\phn & 3.4\phn & 4730\phn & \nodata\phn \\
7\phn & \object[Cl* NGC 2099    HGH V131]{V131}\phn & 055129.26+323235.44\phn & 18.515\phn & 0.005\phn & 0.029\phn & 1.4\phn & 4730\phn & \nodata\phn \\
8\phn & \object[Cl* NGC 2099    HGH V147]{V147}\phn & 055130.48+324053.05\phn & 18.348\phn & 0.005\phn & 0.033\phn & 1.9\phn & 4730\phn & 7.709\phn \\
9\phn & \object[Cl* NGC 2099    HGH V157]{V157}\phn & 055131.24+323059.61\phn & 17.688\phn & 0.008\phn & 0.046\phn & 12.9\phn & 4730\phn & \nodata\phn \\
10\phn & \object[Cl* NGC 2099    HGH V163]{V163}\phn & 055131.87+324233.83\phn & 17.482\phn & 0.004\phn & 0.026\phn & 5.1\phn & 4730\phn & \nodata\phn \\
\enddata
\tablecomments{Table \ref{TabA3} is published in its entirety in the electronic edition of {\it Astronomical Journal}.  A portion is shown here for guidance regarding its form and content.}
\tablenotetext{a}{$\sigma_{m}$ is the measured standard deviation of 3$\sigma$-clipped light curve.}
\tablenotetext{b}{$\Delta m$ is the peak-to-trough amplitude of 3$\sigma$-clipped light curve ($\Delta m$ = $m_\mathrm{max}$ $-$ $m_\mathrm{min}$)}
\tablenotetext{c}{The reduced chi-square $\chi^{2}$ ($\chi^{2}_{\nu}$) of the magnitudes is to test the significance of variability for individual objects against the null hypothesis of no variation.}
\tablenotetext{d}{For the objects that were previously recognized as periodic variables, we fill the column with the period value using the catalog of \citet{har08b}.}
\end{deluxetable}
\clearpage

% Table A4
\begin{deluxetable}{rccrcrrrc}
\tabletypesize{\footnotesize}
\tablecaption{436 flare stars\label{TabA4}}
\tablewidth{0pt}
\tablehead{
\colhead{} & \colhead{} & \colhead{Star} & \colhead{$r$} & \colhead{Flare$_\mathrm{event}$} & \colhead{$t_\mathrm{peak}$\tablenotemark{a}} & \colhead{$\Delta$m$_\mathrm{peak}$\tablenotemark{b}} & \colhead{$\tau_{0.2}$\tablenotemark{c}} & \colhead{}\\
\colhead{No.} & \colhead{VarID} & \colhead{(J2000)} & \colhead{(mag)} & \colhead{(No.)} & \colhead{(days)} & \colhead{(mag)} & \colhead{(hours)} & \colhead{Note\tablenotemark{d}}
}
\startdata
1\phn & \object[Cl* NGC 2099    HGH V42]{V42}\phn & 055120.47+322200.75\phn & 20.690\phn & 1\phn & 53742.40307\phn & 0.093\phn & 0.17\phn & \phn \\ 
 \phn & \phn & \phn & \phn & 2\phn & 53742.46247\phn & 0.135\phn & 0.08\phn & \phn \\ 
2\phn & \object[Cl* NGC 2099    HGH V50]{V50}\phn & 055121.46+322856.90\phn & 19.990\phn & 1\phn & 53730.34740\phn & 0.221\phn & 0.55\phn & \phn \\ 
 \phn & \phn & \phn & \phn & 2\phn & 53733.19093\phn & 0.176\phn & 0.21\phn & \phn \\ 
 \phn & \phn & \phn & \phn & 3\phn & 53735.19767\phn & 0.039\phn & 0.49\phn & \phn \\ 
 \phn & \phn & \phn & \phn & 4\phn & 53737.33376\phn & 0.188\phn & 0.30\phn & \phn \\ 
3\phn & \object[Cl* NGC 2099    HGH V69]{V69}\phn & 055123.45+322948.83\phn & 20.809\phn & 1\phn & 53732.22482\phn & 0.061\phn & 0.32\phn & \phn \\ 
 \phn & \phn & \phn & \phn & 2\phn & 53732.35513\phn & 0.133\phn & 1.11\phn & \phn \\ 
 \phn & \phn & \phn & \phn & 3\phn & 53742.35427\phn & 0.281\phn & 0.23\phn & \phn \\ 
4\phn & \object[Cl* NGC 2099    HGH V72]{V72}\phn & 055124.44+324407.71\phn & 15.143\phn & 1\phn & 53730.32306\phn & 0.010\phn & 1.47\phn & \phn \\ 
5\phn & \object[Cl* NGC 2099    HGH V76]{V76}\phn & 055124.78+323440.56\phn & 18.368\phn & 1\phn & 53728.24480\phn & 0.026\phn & 0.17\phn & \phn \\ 
 \phn & \phn & \phn & \phn & 2\phn & 53742.44376\phn & 0.117\phn & 0.24\phn & \phn \\ 
 \phn & \phn & \phn & \phn & 3\phn & 53753.25302\phn & 0.139\phn & 0.16\phn & \phn \\ 
6\phn & \object[Cl* NGC 2099    HGH V78]{V78}\phn & 055124.82+322937.32\phn & 20.017\phn & 1\phn & 53726.39533\phn & 0.046\phn & 0.10\phn & Var\phn \\ 
7\phn & \object[Cl* NGC 2099    HGH V86]{V86}\phn & 055125.41+323800.03\phn & 18.086\phn & 1\phn & 53726.27563\phn & 0.041\phn & 0.58\phn & Var\phn \\ 
 \phn & \phn & \phn & \phn & 2\phn & 53727.30272\phn & 0.025\phn & 0.46\phn & \phn \\ 
8\phn & \object[Cl* NGC 2099    HGH V91]{V91}\phn & 055125.71+323106.38\phn & 20.553\phn & 1\phn & 53734.23990\phn & 0.072\phn & 0.20\phn & Var\phn \\ 
 \phn & \phn & \phn & \phn & 2\phn & 53756.08268\phn & 0.079\phn & 0.60\phn & \phn \\ 
9\phn & \object[Cl* NGC 2099    HGH V104]{V104}\phn & 055126.78+324024.73\phn & 17.971\phn & 1\phn & 53726.43908\phn & 0.030\phn & 0.29\phn & \phn \\ 
10\phn & \object[Cl* NGC 2099    HGH V112]{V112}\phn & 055127.86+322854.79\phn & 22.181\phn & 1\phn & 53727.19210\phn & 0.360\phn & 0.58\phn & Var\phn \\ 

\enddata
\tablecomments{Table \ref{TabA4} is published in its entirety in the electronic edition of {\it Astronomical Journal}.  A portion is shown here for guidance regarding its form and content.}
\tablenotetext{a}{$t_{peak}$ is the time at flare peak.}
\tablenotetext{b}{$\Delta m_{peak}$ is the incremental magnitude at flare peak.}
\tablenotetext{c}{We define the flare duration  ($\tau_{0.2}$=$t_\mathrm{0.2,rise}$ + $t_\mathrm{0.2,decay}$) as the time-scales of flare events at 80\% of its peak level (Paper III).}
\tablenotetext{d}{These stars show periodic brightness variations with a near or distorted sinusoidal shape (Var).}
\end{deluxetable}
\clearpage


\begin{thebibliography}{}
\bibitem[Abazajian et al.(2009)]{aba09} Abazajian, K. N., Adelman-McCarthy, J. K., Ag\H{u}eros, M. A., et al. 2009, \apjs, 182, 543
\bibitem[Andronov \& Chinarova(2013)]{and13} Andronov, I. L. \& Chinarova, L. L. 2013, arXiv:1308.1129
\bibitem[Balona(2012)]{bal12} Balona, L. A. 2012, \mnras, 423, 3420

\bibitem[Benk\H{o} et al.(2010)]{ben10} Benk\H{o}, J. M., Kolenberg, K., Szab\'{o}, R., et al. 2010, \mnras, 409, 1585
\bibitem[Benz \& G\"{u}del(2010)]{benz10} Benzm, A. O. \& G\"{u}del, M. 2010, \araa, 48, 241
\bibitem[Breger et al.(1993)]{bre93} Breger, M., Stich, J., Garrido, R., et al. 1993, \aap, 271, 482
\bibitem[Butler \& Bloom(2011)]{but11} Butler, N. R. \& Bloom, J. S. 2011, \aj, 141, 93
\bibitem[Chang et al.(2015)]{cha15} Chang, S.-W., Byun, Y.-I., \& Hartman, J. D. 2015, \aj, 149, 135 (Paper I)
\bibitem[Ciardi et al.(2011)]{cia11} Ciardi, D. R., Braun, K. V., Bryden, G., et al. 2011, \aj, 141, 108
\bibitem[Covey et al.(2007)]{cov07} Covey, K. R., Ivezi\'{c}, \v{Z}., Schlegel, D., et al. 2007, \aj, 134, 2398
\bibitem[Cutri et al.(2003)]{cut03} Cutri, R. M., Skrutskie, M. F., van Dyk, S., et al. 2003, The IRSA 2MASS All-Sky Point Source Catalog, NASA/IPAC Infrared Science Archive, \url{http://irsa.ipac.caltech.edu/applications/Gator/}
\bibitem[Debosscher et al.(2007)]{deb07} Debosscher, J., Sarro, L. M., Aert, C., et al. 2007, \aap, 475, 1159
\bibitem[Debosscher et al.(2009)]{deb09} Debosscher, J., Sarro, L. M., L\'{o}pez, M., et al. 2009, \aap, 506, 519
\bibitem[Devor(2005)]{dev05} Devor, J. 2005, \apj, 628, 411
\bibitem[Eyer \& Mowlavi(2008)]{eye08} Eyer, L. \& Mowlavi, N. 2008, J. Phys. Conf. Ser., 118, 012010
\bibitem[Girardi et al.(2004)]{gir04} Girardi, L., Grebel, E. K., Odenkirchen, M., \& Chiosi, C. 2004, \aap, 422, 205
\bibitem[Girardi et al.(2005)]{gir05} Girardi, L., Groenewegen, M. A. T., Hatziminaoglou, E., \& da Costa, L. 2005, \aap, 436, 895
\bibitem[Glicenstein(2001)]{gli01} Glicenstein, J. -F. 2001, in ASP Conf. Ser. 239, Microlensing 2000: A New Era of Microlensing Astrophysics, ed. J. W. Menzies \& Penny D. Sackett. (San Francisco, CA:ASP), 28
\bibitem[Gromadzki et al.(2009)]{gro09} Gromadzki, M., Miko\l{}ajewska, J., Whitelock, P., \& Marang, F. 2009, Acta Astron., 59, 169
\bibitem[Gromadzki et al.(2013)]{gro13} Gromadzki, M., Mikolajewska, J., \& Soszynski, I. 2013, arXiv:1312.6063
\bibitem[Gruberbauer et al.(2007)]{gru07} Grubrbauer, M., Kolenberg, K., Rowe, J. F., et al. 2007, \mnras, 379, 1498
\bibitem[Guggenberger et al.(2012)]{gug12} Guggenberger, E., Kolenberg, K., Nemec, J. M., et al. 2012, \mnras, 424, 649
\bibitem[Hartman et al.(2008a)]{har08a} Hartman, J. D., Gaudi, B. S., Holman, M. J., et al. 2008a, \apj, 675, 1233
\bibitem[Hartman et al.(2008b)]{har08b} Hartman, J. D., Gaudi, B. S., Holman, M. J., et al. 2008b, \apj, 675, 1254
\bibitem[Hartman et al.(2009a)]{har09a} Hartman, J. D., Gaudi, B. S., Pinsonneault, M. H., et al. 2009a, \apj, 691, 342
\bibitem[Hartman et al.(2009b)]{har09b} Hartman, J. D., Gaudi, B. S., Holman, M. J., et al. 2009b, \apj, 695, 336
\bibitem[Henry et al.(2000)]{hen00} Henry, Gregory W., Fekel, Francis C., Henry, Stephen M., \& Hall, Douglas S. 2000, \apjs, 130, 201
\bibitem[Irwin et al.(2007)]{irw07} Irwin, J., Irwin, M., Aigrain, S., et al., \mnras, 375, 1449
\bibitem[Kalirai et al.(2001)]{kal01} Kalirai, J. S., Ventura, P., Richer, H. B., et al. 2001, \aj, 122, 3239
\bibitem[Kang et al.(2007)]{kan07} Kang, Y. B., Kim, S.-L., Rey, S.-C., et al. 2007, \pasp, 119, 239
\bibitem[Kelly et al.(2009)]{kel09} Kelly, B. C., Bechtold, J., \& Siemiginowska, A. 2009, \apj, 698, 895
\bibitem[Kiss et al.(2001)]{kis01} Kiss, L. L., Szab\'{o}, Gy. M., Szil\'{a}di, K., F\"{u}r\'{e}sz, G., S\'{a}rneczky, K., \& Cs\'{a}k, B. 2001, \aap, 376, 561
\bibitem[Kov\'{a}cs et al.(2002)]{kov02} Kov\'{a}cs, G., Zucker, S., \& Mazeh, T. 2002, \aap, 391, 369
\bibitem[Kuschnig et al.(1997)]{kus97} Kuschnig, R., Weiss, W. W., Gruber, R., Bely, P. Y., Jenkner, H. 1997, \aap, 328, 544
\bibitem[Lenz \& Breger(2005)]{len05} Lenz P., Breger M. 2005, CoAst, 146, 53
\bibitem[Lomb(1976)]{lom76} Lomb, N. R. 1976, \apss, 39, 447
\bibitem[Messina et al.(2008)]{mes08} Messina, S., Distefano, E., Parihar, P., et al. 2008, \aap, 483, 253
\bibitem[Molnar et al.(2013)]{mol13} Molnar, Lawrence A., Van Noord, Daniel M., \& Steenwyk, Steven D. 2013, arXiv:1310.0539
\bibitem[Richards et al.(2011)]{ric11} Richards, J. W., Starr, D. L., Butler, N. R., et al. 2011, \apj, 733, 10
\bibitem[Rucinski(1992)]{ruc92} Rucinski, S. M. 1992, \aj, 103, 960
\bibitem[Rucinski(2002)]{ruc02} Rucinski, S. M. 2002, \pasp, 114, 1124
\bibitem[Palmer(2009)]{pal09} Palmer, D. M. 2009, \apj, 695, 496
\bibitem[Plavchan et al.(2008)]{pla08} Plavchan, P., Jura, M., Kirkpatrick, D., Cutri, Roc M., \& Gallagher, S. C. 2008, \apjs, 175, 191
\bibitem[Press \& Rybicki(1989)]{pre89} Press, W.H. \& Rybicki, G.B. 1989, \apj, 338, 277
\bibitem[Press et al.(1992)]{pre92} Press, W. H., Teukolsky, S. A., Vetterling, W. T., \& Flannery, B. P. 1992, Numerical Recipes in C. 2nd ed. (New York: Cambridge University Press)
\bibitem[Reegen(2007)]{ree07} Reegen, P. 2007, \aap, 467, 1353
\bibitem[Samus et al.(2009)]{sam09} Samus, N. N., Durlevich, O. V., Kazarovets, E. V., et al. 2009, VizieR Online Data Catalog:B/gcvs, 1, 2025
\bibitem[Scargle(1982)]{sca82} Scargle, J. D. 1982, \apj, 263, 835
\bibitem[Sch\"{u}tz \& Holschneider(2011)]{sch11} Sch\"{u}tz, N. \& Holschneider, M. 2011, \pre, 84, 021120
\bibitem[Schwarzenberg-Czerny(1989)]{sch89} Schwarzenberg-Czerny, A. 1989, \mnras, 241, 153
\bibitem[Shin \& Byun(2004)]{shi04} Shin, M.-S. \& Byun, Y.-I. 2004, JKAS, 37, 79
\bibitem[Taylor(2000)]{tay00} Taylor, W. 2000, in Change-Point Analyzer 2.0 shareware program, Taylor Enterprises, Libertyville, Illinois., \url{http://www.variation.com/cpa}
\end{thebibliography}
\end{document}